# FDTD-based Inverse Design enables f/0.27 flat Microlens Array for Integral Imaging

Tina M. Hayward, Robert Stewart, Rajesh Menon, and Apratim Majumder

*Abstract*—We demonstrate a high-NA (0.88), ultra-low-f-number (f/0.2714), multi-wavelength (480nm, 550nm and 650nm) multilevel diffractive MicroLens Array (MLA) using inverse design. Each microlens in the array is close-packed with diameter of 70 µm and focal length of only 19 µm in air. The MLA was patterned on one surface of a polymer film via UV casting, such that the focal plane was located on the distal end of the film (n of polymer ~ 1.47, thickness = 28 µm, effective f/# (NA) inside polymer ~ 0.4 (0.78)). Each microlens focuses incident light at 3 design wavelengths into a focal spot with measured full-width at half-maximum (FWHM) < 1 µm. By placing this MLA directly on a high-resolution print, we demonstrated RGB integral imaging with applications in document security. Compared to refractive MLAs, our diffractive MLA reduces the thickness by > 3X, which is advantageous for manufacturability. Since these multi-level diffractive MLAs are fabricated using UV-casting, they have the potential for low-cost, high-volume manufacturing.

## I. INTRODUCTION

IN integral imaging, a microlens array (MLA) is used to capture as well as to reconstruct full light-field information [1], [2], [3], [4], [5], [6], [7], [8], [9]. Light field cameras were typically implemented using refractive MLAs [10], [11], [12], [13], [14]. 3D integral images were achieved by combining MLAs with parallax-integrated offset prints or displays [15], [16], [17], [18]. A printed image (or a display), placed approximately one focal length away from the MLA produces 3D images due to multiple parallaxes when viewed from different angles, see Fig. 1(a). The observed image changes with viewing angle, which creates an impression of depth. Therefore, MLAs have an important role to play in anti-counterfeiting, for example in banknotes [19]. Security increases with complexity of the optical features, number of colors used, and brightness and contrast of the images produced. Larger microlenses allow the incorporation of more views and thereby, more complex 3D images, enhancing security [20]. The thickness of such a security feature is typically limited by the focal length of the microlens. Attaining large-diameter microlens with short focal length, i.e., a small f/# is very challenging due to the typical poor optical performance and manufacturing complexities. Thinner security features and therefore, ultra-low f-number MLAs are highly desirable. Refractive microlenses used in most cases [15], [16], [17], [18], [19], [21], [22], [23], suffer from chromatic aberration, low-NA (large f/#), and thus, low focusing resolution in addition to large device sag and focal length (therefore, thickness). Lastly, it is desirable to attain a large depth-of-focus (DOF) to compensate for the expected thickness variations in the print films (usually 2-3 µm). Some proposed solutions include holographic optical elements, which unfortunately suffer from poor broadband performance [24] and multiple MLAs [25], which increase the thickness of the overall system, and require complicated alignment procedures. Low f/# Fresnel lenses have been demonstrated before, but these suffer from significant chromatic aberrations [26], and relatively low efficiency [27], [28]. We note that metalens-based MLAs have been demonstrated recently, [29], [30], [31], [32] but these generally suffer from poor efficiency and complex fabrication requirements (see Table S1 in supplement and related discussion) [33]. Here, we demonstrate an approach to solving these problems via an inverse-designed multilevel diffractive MLA having diameter of 70 µm, and focal length (in air) of 19 µm that is achromatic over 3 design wavelengths of 480, 550 and 650nm. In addition, we use a high-volume manufacturing process to replicate our MLA.

Multilevel diffractive lenses (MDLs) have been demonstrated for various applications [34], [35], [36], [37], [38], [39], [40], [41], [42], [43]. Previously, we showed that inverse design via PSF engineering can be used to decouple the f/# from the NA, *i.e.*, allow the microlens to form a focal spot larger than the diffraction limit without compromising its light-collection ability [44], [45]. In all prior cases, scalar diffraction coupled with inverse design was used. However, when both ultra-low f/# and diffraction-limited performance are desired, then scalar theory falls short, which we address here. Specifically, we report three advances: (1) inverse design using rigorous finite-difference-time-domain (FDTD) to create microlenses with diameter = 70 µm and focal length (in air) = 19 µm (f/0.2714) as illustrated in Fig. 1(b); (2) master fabrication using grayscale lithography with minimum feature width of 700nm, which is then used for UV-casting replication into the top 1 µm of a ~29 µm-thick polymer film (the backing 28 µm thickness corresponds to focal length of MLA in the

This work was supported in part by the Office of Naval Research under Grant N6560-NV-ONR, and the National Science Foundation under Grant CMMI-222903. *(Corresponding author: Apratim Majumder).*

Apratim Majumder and Tina M. Hayward are with the Department of Electrical and Computer Engineering, University of Utah, Salt Lake City, UT 84112 USA (e-mail: apratim.majumder@utah.edu).

Robert Stewart is with Koenig and Bauer Banknote Solutions, Lausanne, Switzerland, CH1018.

Rajesh Menon is with the Department of Electrical and Computer Engineering, University of Utah, Salt Lake City, UT 84112 USA and Oblate Optics Inc., San Diego, CA 92130 USA.

Supplemental materials and visualization are available at [33].

polymer); and (3) full optical characterization including measurement of PSF and demonstration of integral-imaging under ambient white light, when coupled with an offset print.

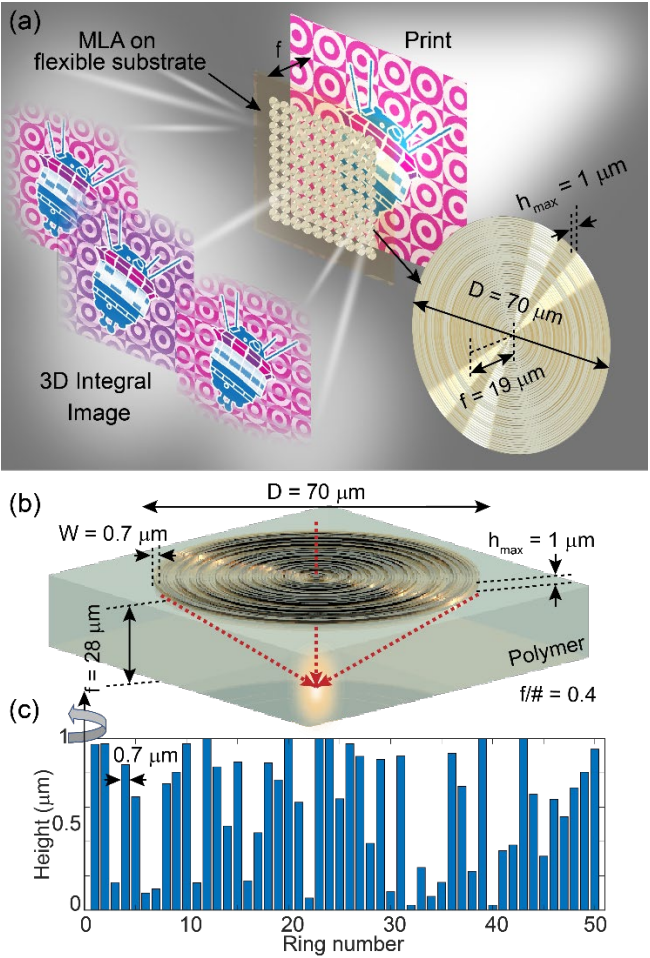

**Fig. 1.** (a) Illustration of 3D integral image generation by combining an MLA (f/0.27 in air) with a high-resolution print (notice the changing pattern at the center of the image based on the viewing angle (also see Visualizations 4 and 5). The MLA is placed ~one focal length, f, away from the print. (b) Each microlens imprinted on the front surface of a transparent polymer film focuses incident collimated light to a near-diffraction limited spot on its back surface (thickness = f = 28 µm). The polymer film has refractive index ~ 1.47, which leads to f/# ~ 0.4 and NA ~ 0.78 for the microlens. The MLA is composed of concentric rings of different heights (< 1 µm) and fixed ring width (W = 0.7 µm). The diameter of each microlens, D = 70 µm (50 rings). (c) Radial distribution of ring heights of one microlens.

## II. INVERSE DESIGN

We implemented inverse design using the FDTD model in cylindrical coordinates to exploit rotational symmetry for computational efficiency. In inverse design, the desired optical function/response is first selected and an algorithm is applied to generate the optical design that can perform such function.

Following this approach, the objective of inverse design, for our case (result in Fig. 1(c) and Fig. 2(c)) is to maximize a figure-of-merit (FOM) Fig. 2(a)), set as the ratio of power in the desired focal volume (defined as a cylinder of diameter of 5 µm x length of 2 µm) to the total power within the concentration volume centered at a nominal focal distance of 19 µm (in air) for 3 design wavelengths (480, 550, 650nm). Then we apply a modified version of the direct binary search (DBS) algorithm to generate the height distribution of the constituent concentric rings of the MDL that achieves this FOM. The electromagnetic fields after the microlens is simulated using FDTD [46] via MEEP, an open-source FDTD solver [47]. Rotational symmetry enables much faster computation using cylindrical coordinates as it reduces the computational domain as illustrated in Fig. 2(a) [48]. Inverse design is performed using the modified binary search as described elsewhere [36], [37]. Details are provided in section 3 of the Supplement [33]. The microlens is discretized into 50 rings each of width, 700 nm. We allowed the ring heights to vary from 0 to 1 µm over 32 levels. The inverse design converges within ~10 iterations, as shown in Fig. 2(b). The optimized design is shown in Fig. 2(c) with the optimized ring heights, shown in Fig. 1(c). It is to be noted that a traditional Fresnel Zone Plate (FZP) lens, with the same diameter and f/# cannot achieve the same performance as it suffers from significant chromatic aberrations. We simulated the performance of an FZP to demonstrate this (Section 5 of the Supplement).

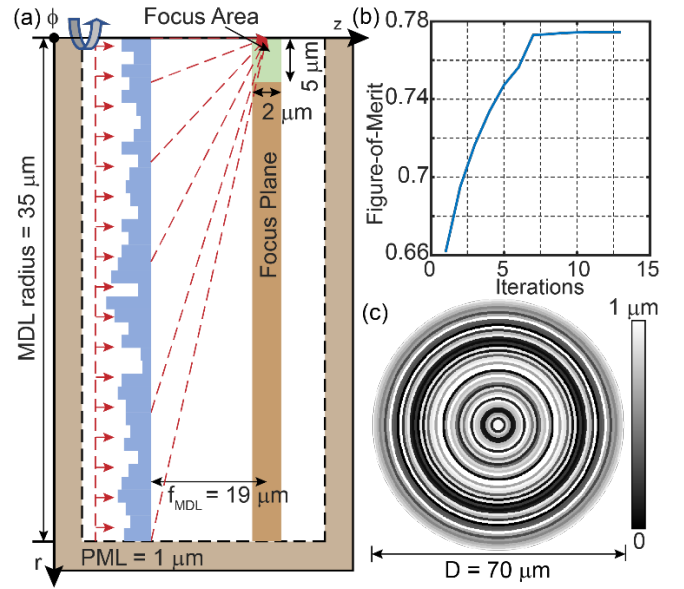

**Fig. 2.** (a) We exploit rotational symmetry to efficiently compute the fields using FDTD. Inverse design maximizes a figure-of-merit (FOM), which is set as the ratio of power inside a desired focal volume near the optical axis (green cylinder with radius=5 µm and thickness=2 µm) to that in the total volume located at the focal plane (also see Supplement section 3) [33]. (b) The FOM converges within ~10 iterations. (c) Top view of the optimized design.

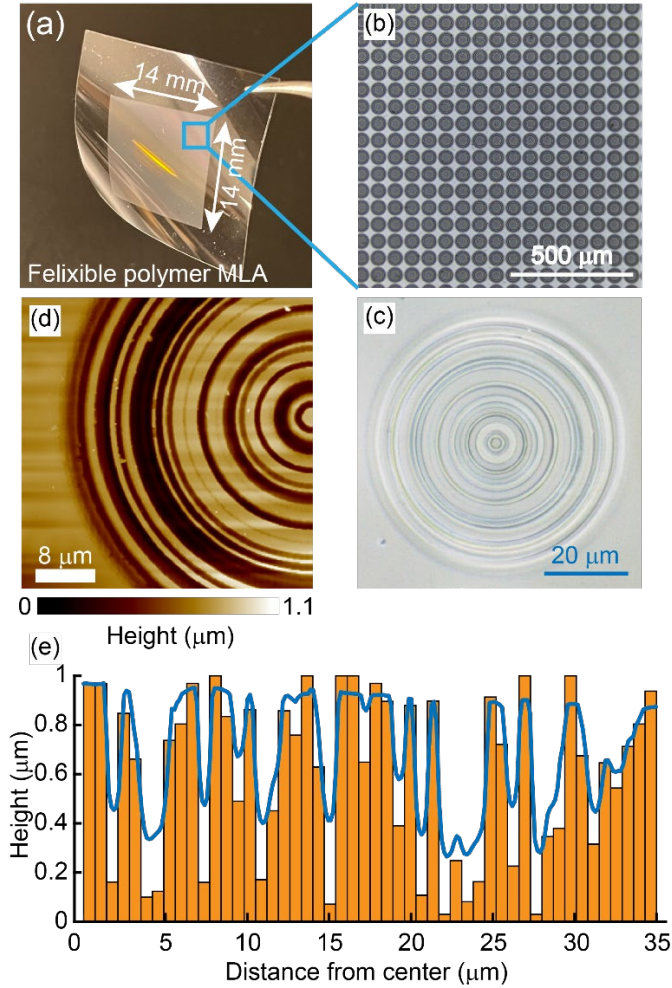

**Fig. 3.** (a) Replicated MLA. (a) Photograph of the MLA film. (b-c) Optical micrographs at different magnifications. (d) Atomic force micrograph (AFM) of one of the μ-MDLs in the array. (e) Measured ring-heights (blue) compared to corresponding design values (orange bars). Black curve denotes the error in ring heights. Mean height error = 111.4 nm, Standard deviation of height error = 94.6 nm.

## III. FABRICATION

We first fabricated a 200×200 μ-MDL array master (size = 14 mm × 14 mm) in photoresist (Shipley 1813, Microchem) on a 50 mm-diameter sodalime glass substrate using grayscale lithography via a laser pattern generator (DWL66+, Heidelberg Instruments GmbH) [49]. The master was replicated using WaveFront Technology Inc.'s UV cast & cure process (see details in supplement section 4) [33]. Fig. 3(a) shows a photograph of the flexible MLA on the polymer film, while optical micrographs of the replicated MLA are shown in Figs. 3(b) and 3(c). The microlenses are close packed with center-to-center spacing of 70μm. The thickness of the polymer film is ~ 28 μm, equal to the focal length of the microlens inside the polymer. The flexibility of the polymer MLA allows it to be readily integrated with high-resolution prints, to produce integral images. The ring heights of one of the microlenses in the array was measured using an atomic force microscope (AFM, Bruker Dimension Icon). The AFM data is shown in Fig. 3(d) along with a plot of the heights compared to the ideal design heights and the error in Fig. 3(e). The error is calculated as Δ*Height = Measured height - ideal height*. The average difference between the measured and designed heights was 111.4 nm, with standard deviation of 94.6 nm. This error is typical of master fabrication as evidenced in our prior works [36], [38], [39], [41], [42], [43], and is acceptable for device functionality. We also simulated the effect of the fabrication errors on the performance of the MLA, exhibiting PSF broadening and higher sidelobes, that is also observed in the experiments (see Section 5 of the Supplement).

## IV. RESULTS AND DISCUSSION

The simulated wavelength-averaged axial point-spread function of the microlens is shown in Fig. 4(a) and 4(c) for air and polymer behind the microlens, respectively. The nominal focal plane (white dashed lines) is located at the peak of the on-axis intensity (red line), 18.51 μm in air and 30.09 μm in polymer, away from the microlens. This is within the 2 μm target DOF from the target of 19 μm in air and ~ 28.5 μm in polymer. The wavelength-averaged point-spread function (PSF) at this plane is plotted in Fig. 4(b) and 4(c) for air and polymer, respectively. A collimated light source with a tunable wavelength filter and a magnification system with a monochromatic CMOS image sensor were used to measure the PSFs. In order to keep the polymer film rigid during measurements, it was mounted on a laser-cut acrylic holder with a window using Kapton tape (see section 5 of Supplement) [33]. The wavelength-averaged measurements are shown in Figs. 4(e) and 4(f). There is some broadening of the PSF, and appearance of sidelobes in the measurements. We attribute this to the rounded corners of the rings resulting from our master fabrication (see Figs. 4g and h, and section 5, Supplement), which is similar to conclusions from other works [50].

Next, we placed the MLA atop offset color prints to test integral imaging (patterned side faced the observer). The prints were generated using standard banknote plate making and offset printing equipment. Original designs were generated at 10,160 dpi (pixel size 2.5 μm), resolution losses in plate making and printing limit the minimum feature size to 10 - 15 μm. The feature designs incorporated both line and icon Moiré effects, which were selected for their tolerance to expected effective resolution reduction caused by the print process, and also for the simplicity of the resulting optical effects (see section 4 of Supplement and Visualizations 4 and 5). We tested the MLA with various print geometries and an example is shown in Fig. 5(a). Exemplary images observed at varying viewing angles and with different illumination conditions such as indoors and outdoors are shown in Figs. 5(c), 5(d), 5(e), and 5(f). The images agree well with simulations (Fig. 5(b)). Additional visualizations (1: print without MLA, 2: print with MLA, 3: print with MLA viewed in transmission and reflection, 6: print with MLA held together using Kapton tape) are also included.

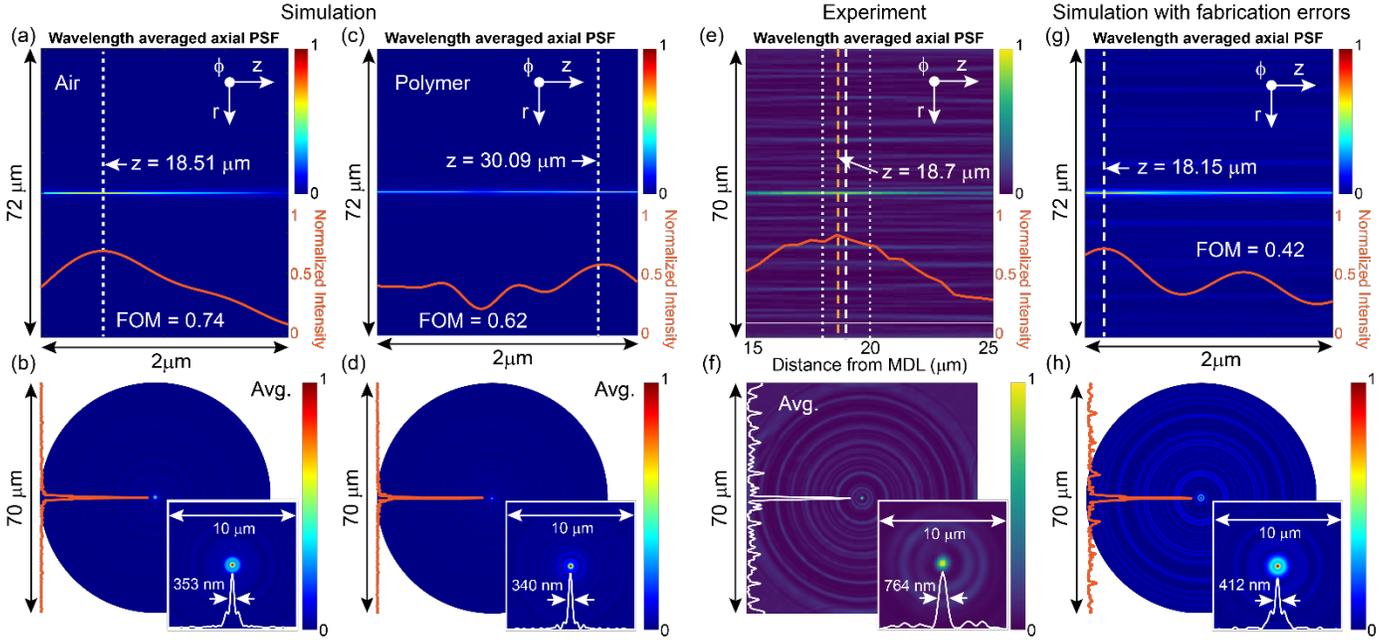

**Fig. 4.** Focusing performance. (a) Simulated wavelength-averaged axial PSF in air and (b) polymer and (e) measured wavelength-averaged axial PSF. Simulated wavelength-averaged transverse PSF (b) in air and (d) in polymer, and (f) corresponding measurement in air (focal plane located at yellow dashed line at 18.7 µm, desired focal plane marked by white dashed line, and desired focal volume marked by dotted white lines). Magnified views (10 µm × 10 µm) and FWHM are shown in the corresponding insets. (g, h) Simulations, same as (a-d), but taking into account errors in fabrication, using height profile data shown in Fig. 3(e), showing presence of sidelobes as observed in the experiments.

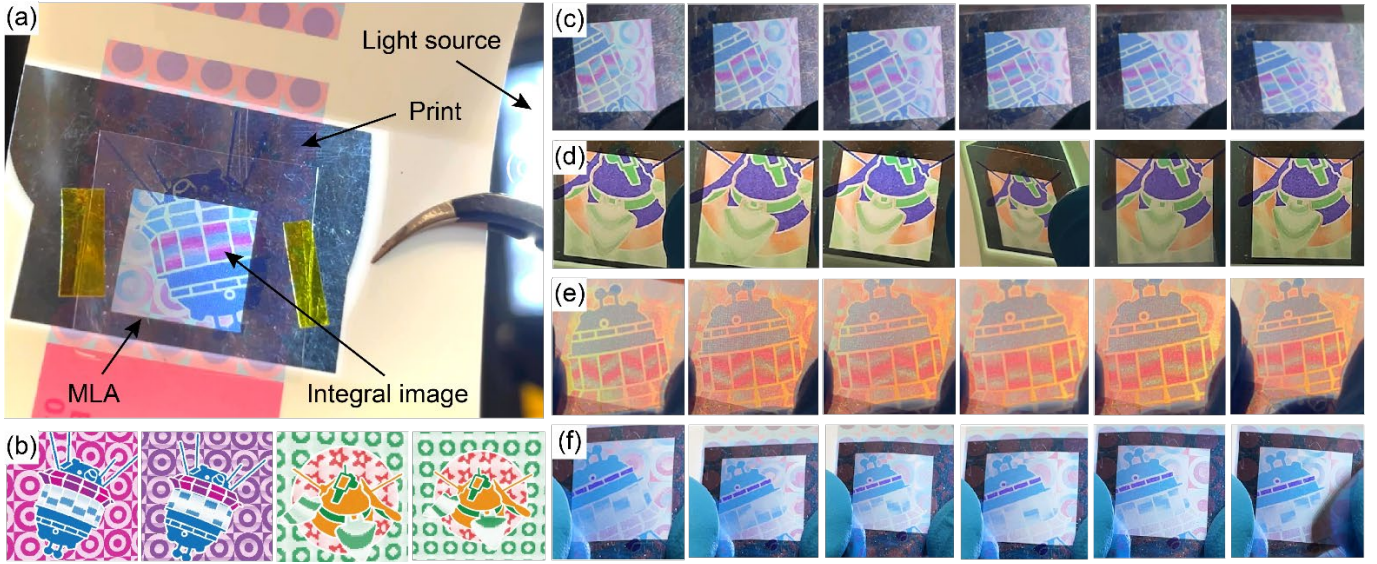

**Fig. 5.** Summary of integral imaging. (a) Photograph of the MLA on offset print held together using Kapton tape, exhibiting integral imaging. (b) Simulated images at two viewing angles from two exemplary prints. (c-f) Images at different viewing angles for two exemplary prints (also see Visualizations 2, 3, and 6). The illuminations used were (c, f) an LED flashlight, (d) fluorescent ceiling lights and (e) ambient sunlight, captured outside around 5 pm, on November 08, 2023 at Salt Lake City, Utah, USA.

Using the same rigorous method, we simulated the performance of a Fresnel Zone Plate (FZP) with diameter = 70 µm, focal length = 19 µm (air), and wavelength = 500 nm (center of our three design wavelengths). As expected, the FZP exhibits strong chromatic aberration and therefore, cannot be used for our application (see Section 5 of the Supplement).

## V. CONCLUSION

We demonstrated the inverse design of an ultra-low f/# (0.27) 3-wavelength micro-MDL array, using a rotationally-symmetric FDTD model. The extremely short focal length, while preserving a large aperture allows easy integration onto security offset prints. The measured FWHM of the PSFs were

< 1 µm and readily produces high-quality integral images with print features of width 10 - 15 µm. We fabricated a master pattern (min. feature = 700 nm) using grayscale lithography, which was then replicated using UV casting (a high-volume, high precision manufacturing process) into a 28 µm thick polymer film. Using simulations, we show that our design outperforms a conventional FZP. Our demonstration lays the framework for diffractive optics to attain very low f/# imaging. Thus, in summary, the key contributions of this work are: (1) inverse design of flat multi-level diffractive microlenses with diameter = 70 µm and focal length (in air) = 19 µm (f/0.2714) using FDTD and DBS optimization, (2) master fabrication using grayscale lithography and replication into flexible polymer film using UV-casting, and (3) full optical characterization including measurement of PSF and demonstration of integral-imaging under ambient white light, when coupled with an offset print.

### ACKNOWLEDGMENT

We thank Brian Baker and Joe Spencer (Utah nanofab), and Chris Kent and Kent Coulter (Wavefront Technologies) for assistance with fabrication.

### DISCLOSURE

Rajesh Menon: Oblate Optics (I, E, P).

### REFERENCES

[1] G. Lippmann, "Épreuves réversibles donnant la sensation du relief," J. Phys. Theor. Appl., **7**(1), 821-825 (1908).

[2] J. Arai, H. Kawai, and F. Okano, "Microlens arrays for integral imaging system," Appl. Opt. 45, 9066-9078 (2006).

[3] Q. Xu, B. Dai, Z. Jiao, R. Hong, Z. Yang, D. Zhang, and S. Zhuang, "Fabrication of large micro-structured high-numerical-aperture optofluidic compound eyes with tunable angle of view," Opt. Exp. 26, 33356-33365 (2018).

[4] Z. Zang, X. Tang, X. Liu, X. Lei, and W. Chen, "Fabrication of high quality and low cost microlenses on a glass substrate by direct printing technique," Appl. Opt. 53, 7868-7871 (2014).

[5] S. Zhang, L. Zhou, C. Xue, and L. Wang, "Design and simulation of a superposition compound eye system based on hybrid diffractive-refractive lenses," Appl. Opt. 56, 7442-7449 (2017).

[6] K. H. Jeong, J. J. Kim, and L. P. Lee, "Biologically inspired artificial compound eyes," Science, vol. 312, pp. 557-561 (2006).

[7] H. Jung and K.-H. Jeong, "Monolithic polymer microlens arrays with high numerical aperture and high packing density", ACS Applied Materials & Interfaces, vol. 7, pp. 2160-2165 (2015).

[8] A. Lumsdaine, and T. Georgiev, "Full resolution lightfield rendering," Adobe Technical Report, Adobe Systems, 1-12 (2008).

[9] M. Xu, Y. Xue, J. Li, L. Zhang, H. Lu, and Z. Wang, "Large-Area and Rapid Fabrication of a Microlens Array on a Flexible Substrate for an Integral Imaging 3D Display," ACS Appl. Mater. Interfaces, 15 (7) 10219–10227 (2023).

[10] T. Adelson, and J. Y. A. Wang, "Single lens stereo with a plenoptic camera," IEEE Transactions on Pattern Analysis and Machine Intelligence 14, 2, 99–106 (1992).

[11] R. Ng, et al., "Light field photography with a hand-held plenoptic Camera," Stanford Tech Report CTSR 2005-02 (2005).

[12] T. Georgiev and A. Lumsdaine, "Reducing plenoptic camera artifacts," Computer Graphics Forum, 29, 1955-1968 (2010).

[13] N. Zeller, F. Quint, and U. Stilla, "Depth estimation and camera calibration of a focused plenoptic camera for visual odometry," ISPRS J. Photogramm. Remote Sens. 118, 83–100 (2016).

[14] Y. Bok, H. G. Jeon, and I. S. Kweon, "Geometric calibration of micro-lens-based light field cameras using line features," IEEE Trans. Pattern Anal. Mach. Intell. 39, 287–300 (2017).

[15] X. Zhou, Y. Peng, R. Peng, X. Zeng, Y. Zhang and T. Guo, "Fabrication of large-scale microlens arrays based on screen printing for integral imaging 3D display," ACS Appl. Mater. Interfaces 8(36), 24248–24255 (2016).

[16] S. Li, Q.-H. Wang, Y.-P. Xia, Y. Xing, H. Ren, H. Deng, "Integral imaging 3D display system with improved depth of field using a colloidal scattering layer," Optics Communications, vol. 484 (2021).

[17] Q. Wang, H. Deng, T. Jiao, D. Li, and F. Wang, "Imitating micro-lens array for integral imaging," Chin. Opt. Lett. 8, 512-514 (2010).

[18] Y. Xing, et al., "Integral imaging-based tabletop light field 3D display with large viewing angle," Opto-Electronic Advances, 6 (6), 220178 (2023).

[19] https://www.uscurrency.gov/denominations/100

[20] M. Dejean, V. Nourrit and J.-L. de Bougrenet de la Tocnaye, "Object kinetics perception in auto-stereoscopic vision," J. Opt. Soc. Am. A 36(11) C104-C112 (2019).

[21] A. Karimzadeh, "Integral imaging system optical design with aberration consideration," Appl. Opt. 54(7), 1765–1769 (2015)

[22] S. Lee, C. Jang, J. Cho, J. Yeom, and B. Lee, "Viewing angle enhancement of an integral imaging display using Bragg mismatched reconstruction of holographic optical elements," Appl. Opt. 55(3), A95–A103 (2016).

[23] W. Wang, G. Chen, Y. Weng, et al., "Large-scale microlens arrays on flexible substrate with improved numerical aperture for curved integral imaging 3D display," Sci. Rep. 10, 11741 (2020).

[24] S. Lee, C. Jang, J. Cho, J. Yeom, and B. Lee, "Viewing angle enhancement of an integral imaging display using Bragg mismatched reconstruction of holographic optical elements," Appl. Opt. 55(3), A95–A103 (2016).

[25] A. Karimzadeh, "Integral imaging system optical design with aberration consideration," Appl. Opt. 54(7), 1765–1769 (2015).

[26] M. Moghimi, J. Fernandes, A. Kanhere, et al., "Micro-fresnel-zone-plate array on flexible substrate for large field-of-view and focus scanning," Sci. Rep., 5, 15861 (2015).

[27] D. Chao, A. Patel, T. Barwicz, H. I. Smith, and R. Menon, "Immersion Zone-Plate-Array Lithography," J. Vac. Sci. Technol. B, 23(6), 2657-2661 (2005).

[28] D. W. Prather, D. Pustai, and S. Shi, "Performance of multilevel diffractive lenses as a function of f-number," Appl. Opt. 40, 207-210 (2001).

[29] R. J. Lin, V. C. Su, S. Wang, et al., "Achromatic metalens array for full-colour light-field imaging," Nat. Nanotechnol. 14, 227–231 (2019).

[30] Z. B. Fan, H. Y. Qiu, H. L. Zhang, et al., "A broadband achromatic metalens array for integral imaging in the visible," Light Sci. Appl. 8, 67 (2019).

[31] Z. Yang, Z. Wang, Y. Wang, et al., "Generalized Hartmann-Shack array of dielectric metalens sub-arrays for polarimetric beam profiling," Nat. Commun. 9, 4607 (2018).

[32] T. Hu, Q. Zhong, N. Li, Y. Dong, Z. Xu, D. Li, Y. H. Fu, Y. Zhou, K. H. Lai, V. Bliznetsov, H. Lee, W. L. Loh, S. Zhu, Q. Lin, and N. Singh, "A metalens array on a 12-inch glass wafer for optical dot projection," Opt. Fiber Commun. Conf. (OFC) 2020, OSA Technical Digest (Optical Society of America, 2020), paper W4C.3. (2020).

[33] Supplementary Information.

[34] S. Banerji, M. Meem, A. Majumder, F. Guevara Vasquez, B. Sensale-Rodriguez, and R. Menon, "Imaging with flat optics: metalenses or diffractive lenses?," Optica 6, 805-810 (2019).

[35] S. Banerji, M. Meem, A. Majumder, B. Sensale-Rodriguez and R. Menon, "Diffractive flat lens enables extreme depth-of-focus imaging," Optica 7(3), 214-2017 (2020).

[36] P. Wang, N. Mohammad and R. Menon, "Chromatic-aberration-corrected diffractive lenses for ultra-broadband focusing," Sci. Rep., 6, 21545 (2016).

[37] S. Banerji and B. Sensale-Rodriguez, "A computational design framework for efficient, fabrication error-tolerant, planar THz diffractive optical elements," Sci. Rep., 9, 5801 (2019).

[38] M. Meem, S. Banerji, A. Majumder, C. Pies, T. Oberbiermann, B. Sensale-Rodriguez & R. Menon, "Inverse-designed flat lens for imaging in the visible & near-infrared with diameter > 3mm and NA=0.3," Appl. Phys. Lett. 117(4), 041101 (2020).

[39] S. Banerji, M. Meem, A. Majumder, F. Guevara Vasquez, B. Sensale-Rodriguez and R. Menon, "Ultra-thin near infrared camera enabled by a flat multi-level diffractive lens," Opt. Lett. 44(22), 5450-5452 (2019).

[40] M. Meem, S. Banerji, A. Majumder, C. Pies, T. Oberbiermann, B. Sensale-Rodriguez and R. Menon, "Large-area, high-NA multi-level diffractive lens via inverse design," Optica 7(3), 252-253 (2020).

[41] M. Meem, S. Banerji, A. Majumder, F. Guevara Vasquez, B. Sensale-Rodriguez and R. Menon, "Broadband lightweight flat lenses for

longwave-infrared imaging," Proc. Natl. Acad. Sci. 116 (43), 21375-21378 (2019).

[42] D. Lin, T. M. Hayward, W. Jia, A. Majumder, B. Sensale-Rodriguez, and R. Menon, "Inverse-designed multi-level diffractive doublet for wide field-of-view imaging," ACS Photonics 10, 8, 2661–2669 (2023).

[43] S. Banerji, M. Meem, A. Majumder, B. Sensale-Rodriguez, and R. Menon, "Super-resolution imaging with an achromatic multi-level diffractive microlens array," Opt. Lett. 45, 6158-6161 (2020).

[44] M. Meem, A. Majumder, and R. Menon, "Freeform broadband flat lenses for visible imaging," OSA Continuum 4, 491-497 (2021).

[45] A. Majumder, M. Meem, R. Stewart, and R. Menon, "Broadband point-spread function engineering via a freeform diffractive microlens array," Opt. Express 30, 1967-1975 (2022).

[46] A. Taflove and S.C. Hagness, "Computational Electrodynamics: The Finite-Difference Time-Domain Method", Artech: Norwood, 3rd Edition, MA, 2005.

[47] A. Oskooi, D. Roundy, M. Ibanescu, P. Bermel, J.D. Joannopoulos, and S.G. Johnson, "MEEP: A flexible free-software package for electromagnetic simulations by the FDTD method," Computer Physics Communications, Vol. 181, pp. 687-702 (2010).

[48] MEEP Cylindrical coordinates tutorial: https://meep.readthedocs.io/en/latest/Scheme_Tutorials/Cylindrical_Coordinates/#:~:text=Meep%20supports%20the%20simulation%20of,if%20th ere%20is%20sufficient%20symmetry.

[49] Datasheet for DWL66+, Heidelberg Instruments GmbH: https://heidelberg-instruments.com/wp-content/uploads/2021/02/Fact-Sheet-DWL-66-v220408.pdf

[50] C-F. Pan et al., "3D-printed multilayer structures for high–numerical aperture achromatic metalenses," Sci. Adv. 9, eadj9262 (2023).

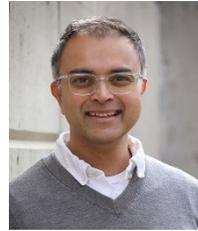

**Rajesh Menon** combines his expertise in nanofabrication, computation and optical engineering to impact myriad fields including super-resolution lithography, metamaterials, broadband diffractive optics, integrated photonics, photovoltaics and computational optics. His research has spawned over 120 publications, over 40 patents, and 4 spin-off companies. Rajesh is a Fellow of the Optical Society of America, and Senior Member of the IEEE and the SPIE. Among his other honors are a NASA Early Stage Innovations Award, NSF CAREER Award and the International Commission for Optics Prize. He currently directs the Laboratory for Optical Nanotechnologies (http://lons.utah.edu/) at the University of Utah. He received S.M. and Ph.D. degrees from MIT.

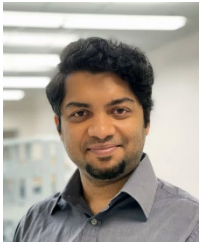

**Apratim Majumder** received his Ph.D. degree in Electrical and Computer Engineering at the University of Utah, where he continues as a research assistant Professor. His research interests cover optics, photonics, hyperspectral imaging, flat optics, and photolithography. He is currently developing ultra-flat lenses and holograms for wide-ranging applications in photography, microscopy, VR/AR, and snapshot hyperspectral cameras for low-cost HIS applications.

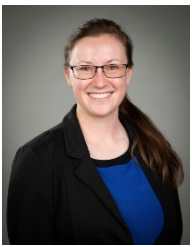

**Tina Hayward** graduated from the University of Utah with a B.S. in Electrical Engineering, and is currently pursuing her Ph.D. in Electrical Engineering. She joined Menon Group in Fall, 2021. Her research interests include grayscale lithography fabrication of multi-level diffractive lenses for extended depth of focus and wide-field of view applications.

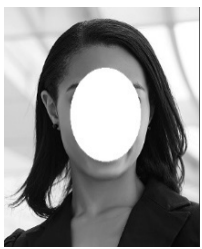

**Robert Stewart** …

# Supplementary Information

# FDTD-based Inverse Design enables f/0.27 flat Microlens Array for Integral Imaging

Tina M. Hayward, Robert Stewart, Rajesh Menon, and Apratim Majumder

## I. LITERATURE SURVEY

Table S1 shows a comparison of this work with those reported in the literature and hence places this work in its appropriate context. For a fair comparison, only flat lenses (Fresnel lenses are excluded) achromatic over a broadband illumination range, or designed for multiple wavelength performance are considered here. Narrowband or single wavelength flat lenses have been ignored. The table has been arranged in ascending order of f/# of the lenses.

**Table S1:** Literature survey of achromatic broadband flat lenses.

| References | Material | Wavelength Range | Bandwidth | Focal Length | Diameter | f/# | Type |
|---|---|---|---|---|---|---|---|
| This work | Polymer | 480, 550, 650 nm | 170 nm | 19 μm (air) 28.5 μm (polymer) | 70 μm | 0.2714 (air) 0.4 (polymer) | Array |
| [1] | a-Si | 1.2 μm – 1.4 μm | 200 nm | 30 μm | 100 μm | 0.3 | Single Lens |
| [2] | PbTe | 5.11 μm – 5.29 μm | 180 nm | 0.5 mm | 1 mm | 0.5 | Single Lens |
| [3] | GaSb | 3 μm – 5 μm | 2 μm | 155 μm | 300 μm | 0.516667 | Single Lens |
| [4] | Polymer | 450 nm – 750 nm | 300 nm | 40 μm | 69 μm | 0.57971 | Array |
| [5] | Au | 532 nm – 1080 nm | 548 nm | 7 μm | 10 μm | 0.7 | Single Lens |
| [6] | a-Si | 3.7 μm – 4.2 μm | 0.5 μm | 300 μm | 300 μm | 1 | Single Lens |
| [7] | Silicon Nitride | 400 nm – 700 nm | 300 nm | 200 μm | 200 μm | 1 | Single Lens |
| [8] | Photoresist on Si | 8 μm – 12 μm | 4 μm | 8 mm | 8 mm | 1 | Single Lens |
| [8] | Photoresist on Si | 8 μm – 12 μm | 4 μm | 19 mm | 15.2 mm | 1.25 | Single Lens |
| [9] | a-Si | 470 nm – 658 nm | 188 nm | 400 μm | 300 μm | 1.333333 | Single Lens |
| [10] | Photoresist | 450 nm – 1000 nm | 550 nm | 5 mm | 3 mm | 1.666667 | Single Lens |

| [11] | Au/SiO2/Au | 1.2 μm – 1.68 μm | 480 nm | 100 μm | 55.55 μm | 1.80018 | Single Lens |
|---|---|---|---|---|---|---|---|
| [12] | IP-Dip photoresist | 1000 nm – 1800 nm | 800 nm | 37.04 μm | 20 μm | 1.852 | Single Lens |
| [1] | a-Si | 1.3 μm – 1.65 μm | 350 nm | 200 μm | 100 μm | 2 | Single Lens |
| [13] | GaN | 400 nm -660 nm | 260 nm | 49 μm | 21.65 μm | 2.263279 | Array |
| [14] | TiO₂ | 490 nm – 550 nm | 60 nm | 485 μm | 200 μm | 2.425 | Single Lens |
| [15] | TiO₂ | 470 nm – 670 nm | 200 nm | 63 μm | 25.2 μm | 2.5 | Single Lens |
| [16] | TiO₂ | 460 nm – 660 nm | 260 nm | 67 μm | 26.4 μm | 2.537879 | Single Lens |
| [17] | Photoresist | 875 nm – 1675 nm | 800 nm | 25 mm | 8.93 mm | 2.799552 | Single Lens |
| [18] | GaN | 435 nm – 685 nm | 250 nm | 20 μm | 7 μm | 2.857143 | Single Lens |
| [19] | TiO₂ | 640 nm – 1200 nm | 560 nm | 35 μm | 10 μm | 3.5 | Single Lens |
| [1] | a-Si | 1.2 μm – 1.65 μm | 450 nm | 800 μm | 200 μm | 4 | Single Lens |
| [12] | IP-Dip photoresist | 1000 nm – 1800 nm | 800 nm | 181.82 μm | 40 μm | 4.5455 | Single Lens |
| [20] | GaN | 400 nm – 660 nm | 260 nm | 235 μm | 50 μm | 4.7 | Single Lens |
| [21] | Fused Si | 486 nm – 656 nm | 170 nm | 100 mm | 20 mm | 5 | Single Lens |
| [19] | TiO₂ | 640 nm – 1200 nm | 560 nm | 75 μm | 15 μm | 5 | Single Lens |
| [22] | Silicon Nitride | 430 nm – 780 nm | 350 nm | 81.5 μm | 14 μm | 5.821429 | Array |
| [19] | TiO₂ | 640 nm – 1200 nm | 560 nm | 150 μm | 20 μm | 7.5 | Single Lens |
| [12] | IP-Dip photoresist | 1000 nm – 1800 nm | 800 nm | 2500 | 300 | 8.333333 | Single Lens |
| [23] | Photoresist | 450 nm – 650 nm | 200 nm | 1 mm | 120 μm | 8.333333 | Array |
| [24] | Photoresist | 450 nm – 750 nm | 300 nm | 45 mm | 4 mm | 11.25 | Single Lens |
| [24] | Photoresist | 450 nm – 750 nm | 300 nm | 45 mm | 4 mm | 11.25 | Single Lens |
| [24] | Photoresist | 450 nm – 750 nm | 300 nm | 45 mm | 4 mm | 11.25 | Single Lens |
| [25] | a-Si | 1300 nm – 1800 nm | 500 nm | 7.5 mm | 600 μm | 12.5 | Single Lens |

Table S2 shows a comparison of this work with other flat lens related works published from our group and thus, places the novelty of this work in its proper context.

**Table S2:** Summary of flat lenses demonstrated by our group in chronological order in the past.

| Ref. | Material | Wavelength Range | Bandwidth | Focal Length | Diameter | f/# | NA | Type |
|---|---|---|---|---|---|---|---|---|
| [26] | Photoresist | 450 – 750 nm | 300 nm | 1 mm | 0.37 mm | 2.7 | 0.18 | Single lens |
| [26] | Photoresist | 450 -750 nm | 300 nm | 1 mm | 0.1 mm | 10 | 0.05 | Single lens |
| [27] | Photoresist | 430 – 660 nm | 230 nm | 25 mm | 2.5 mm | 10 | 0.05 | Single lens |
| [27] | Photoresist | 430 - 660 nm | 230 nm | 25 mm | 2.5 mm | 10 | 0.05 | Single lens |
| [8] | Photoresist on Si | 8 µm – 12 µm | 4 µm | 8 mm | 8 mm | 1 | 0.45 | Single lens |
| [8] | Photoresist on Si | 8 µm – 12 µm | 4 µm | 19 mm | 15.2 mm | 1.25 | 0.37 | Single lens |
| [17] | Photoresist | 875 nm – 1675 nm | 800 nm | 25 mm | 8.93 mm | 2.8 | 0.176 | Single lens |
| [28] | Photoresist | 850 nm | 1 nm | 1 mm | 0.15 mm | 6.67 | 0.075 | Single lens |
| [29] | Photoresist | 850 nm | 35 nm | 1 mm | 4.13 mm | 0.24 | 0.9 | Single lens |
| [10] | Photoresist | 450 nm – 1000 nm | 550 nm | 5 mm | 3 mm | 1.67 | 0.2873 | Single lens |
| [23] | Photoresist | 450 – 650 nm | 200 nm | 1 mm | 0.12 mm | 8.33 | 0.06 | Array |
| [24] | Photoresist | 450 nm – 750 nm | 300 nm | 45 mm | 4 mm | 11.25 | 0.00075, 0.0067, and 0.054 | Single lens |
| [30] | Photoresist | 450 nm - 15 µm | 14.55 µm | 18 mm | 1 mm | 18 | 0.0278 | Single lens |
| [4] | Polymer | 450 nm – 750 nm | 300 nm | 40 µm | 69 µm | 0.39 (air) 0.58 (polymer) | Structured PSF | Array |
| [31] | Photoresist | 4 µm | 1 nm | 25 mm | 25 mm | 1 | 0.45 | Single lens |
| [32] | Photoresist | 850 nm | 34 nm | 1 mm | 0.2 mm | 5 | 0.1 | Doublet for 110 deg FOV |
| This work | Polymer | 480, 550, 650 nm | 170 nm | 19 µm (air) 28.5 µm (polymer) | 70 µm | 0.2714 (air) 0.4 (polymer) | 0.88 | Array |

Table S3 summarizes the features of the above works and provides a comparison of the properties of the device reported in this work, compared to the others.

**Table S3:** Summary of features of flat lenses demonstrated by our group.

| References | Broadband | Very low f/# (≤f/1) | High NA (>0.75) | Extended DOF | Mass replication |
|---|---|---|---|---|---|
| [26] | ✔ | | | | |
| [27] | ✔ | | | | |
| [8] | ✔ | ✔ | | | |
| [17] | ✔ | | | | |
| [28] | | | | | |
| [29] | | ✔ | ✔ | | |
| [10] | ✔ | | | | |
| [23] | ✔ | | | | |
| [24] | ✔ | | | | |
| [30] | ✔ | | | | |
| [4] | ✔ | ✔ | | ✔ | ✔ |
| [31] | | ✔ | | | |
| [32] | | | | | |
| This work | ✔ | ✔ | ✔ | ✔ | ✔ |

## II. MATERIAL PROPERTIES

The dispersion of the nanoimprint polymer material is shown in Fig. S1.

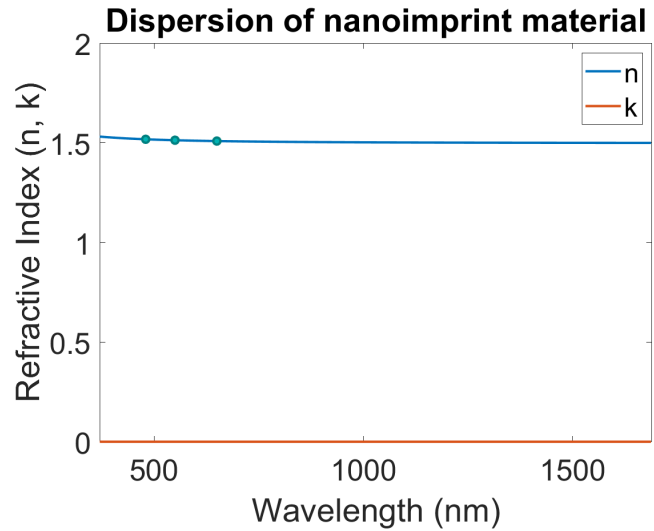

**Fig. S1**: Dispersion (n, k) of the nanoimprint polymer material. The dots represent values for wavelengths 480, 550 and 650 nm.

## III. DETAILS OF INVERSE DESIGN

The objective of the inverse design is to optimize the micro multi-level-diffractive lens (micro-MDL) geometry (with rotational symmetry) that maximizes concentration of normally incident plane wave into a desired focal volume for the 3 design wavelengths. The electromagnetic field after transmitting through the microlens is simulated using FDTD [33] via MEEP (an open-source solver) [34]. Rotational symmetry enables much faster computation using cylindrical coordinates [35]. Fig. S2 shows the computational cell. Since the structure is rotationally symmetric, we needed to simulate only half of the radial cross-section of the micro-MDL. Here, the field vectors are interpreted as $(r, \varphi, z)$ triplets instead of $(x, y, z)$. The radius of the micro-MDL is 35 µm. The computational cell is surrounded on all sides by a perfectly matched layer (PML) of thickness 1 µm. There is a 2 µm air padding preceding the micro-MDL. During optimization, the FDTD simulation is performed using an air layer of thickness 30 µm behind the lens. The center of the focal volume is set at 19 µm in air, which corresponds to $\sim 28.5$ µm in the polymer material. The focal volume is shown by the green box and is a cylinder of diameter = 5 µm and height = 2 µm. The source is set at 0.2 µm away from the boundary of the PML. Thus, the 2D computational cell consists of the $rz$ plane ($\varphi = 0$). The $\varphi$ component of the vector is ignored. In such a cylindrical simulation, the fields can be written in the form of a function of $(r, z)$ multiplied by a function $e^{im\varphi}$, which thus takes into account the $\emptyset$ dependence, and where $m$ is a parameter related to the angular momentum of the field [28].

The figure of merit (FOM) was defined as the wavelength-averaged concentration factor of light, defined as the ratio of power in the desired focal volume to the total power within the concentration volume. The FOM was calculated according to Equation 1 and used during the inverse design optimization process. The optimization was performed running MEEP parallelly across four CPU cores.

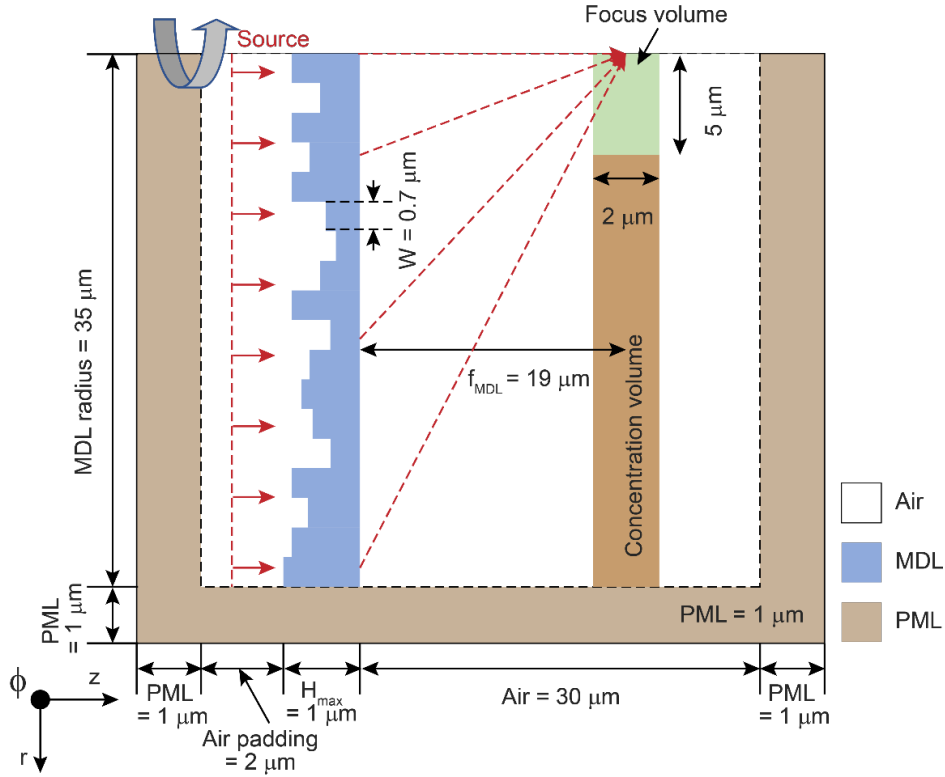

**Fig. S2:** Schematic diagram of the computational cell setup in the open-source FDTD solver MEEP.

$$FOM = \frac{1}{N}\sum_{i=1}^{N} \frac{\iiint_{Focal\ volume} \frac{\epsilon_0}{2}\left|\boldsymbol{E}(\rho,\varphi,z)_{\lambda_i}\right|^2 \cdot d\rho d\varphi dz}{\iiint_{Concentration\ volume} \frac{\epsilon_0}{2}\left|\boldsymbol{E}(\rho,\varphi,z)_{\lambda_i}\right|^2 \cdot d\rho d\varphi dz} \qquad \dots (1)$$

where N = number of design wavelengths, equal to three in our case, corresponding to 480, 550 and 650 nm.

In order to incorporate fabrication constraints, we discretized the radius of the micro-MDL into rings of width, W = 700 nm as shown in Fig. S2. Thus, the total number of rings in one micro-MDL = 50. We allowed the ring heights to vary in discrete steps from 0 to 1 μm over a total of 32 levels. During simulation and optimization, we used a planewave as input to the model, at normal incidence. A linearly-polarized ($x$) planewave, propagating in the $z$ direction is the sum of two circularly-polarized planewaves of opposite chirality [35] as described by Equation 2, as follows:

$$\hat{E}_x = \frac{1}{2}\left[e^{i\phi}(\hat{E}_\rho + i\hat{E}_\phi) + e^{-i\phi}(\hat{E}_\rho - i\hat{E}_\phi)\right] \quad \dots (2)$$

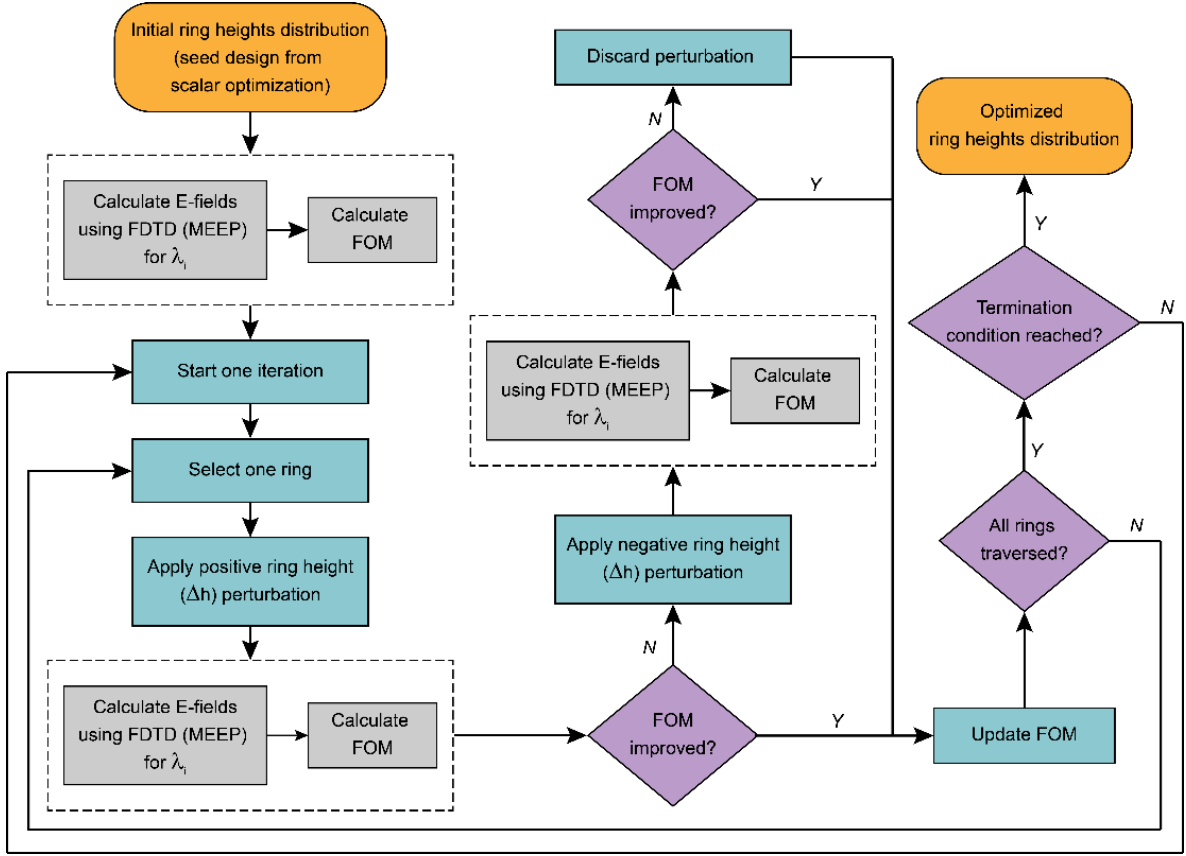

**Fig. S3**: Flow chart of the Direct Binary Search (DBS) algorithm used for optimization. Here, the initial design is the result of another pre-optimization step using the same DBS algorithm shown here, but using scalar theory and MATLAB to calculate the fields. In the pre-optimization step, the initial design is chosen to be a random distribution of ring heights.

The inverse design is performed using the modified direct binary search routine as described elsewhere [36, 37]. In this work, we followed the inverse design protocol in Ref. 36. A summary flow chart of the optimization procedure is shown in Fig. S3. Since the FDTD calculations are still computationally intensive, we opted to use a pre-optimized design as the starting point. The pre-optimization routine follows the same optimization flow chart as shown in Fig. S3, but uses scalar theory and MATLAB to simulate the fields. The starting point for the pre-optimization routine was a random height distribution. As noted previously, scalar theory is not accurate for designing such low f/# lenses and we verified this. Nevertheless, the pre-optimization provides a good starting guess for the actual optimization routine that couples the DBS code to FDTD simulations in MEEP. The results of the optimization including the evolution of the FOM and the final optimized design are shown in Fig. 1(c) and Fig. 1(f), respectively.

Optimization result is shown in Fig. S4. The axial PSF in the entire simulation cell is shown in Fig. S4 (a) for all three design wavelengths (480, 550 and 650 nm). Although half the cell is simulated due to rotational symmetry, for easier data visualization, the other half is concatenated. Fig. S4 (b) shows the wavelength averaged axial PSF in the concentration volume, showing successful convergence of the optimizer. The normalized intensity through the center of the axial PSF is overlaid. The peak corresponds to an on-axis location = 18.46 μm away from the micro-MDL. Fig. S4 (c) shows the magnified view of the axial PSF for each individual design wavelength in the focal volume. The normalized intensity through the center of the axial PSFs for the individual wavelengths as well as the average is plotted in Fig. S4 (d). Next, we extracted the individual cross-sectional PSFs (Normalized intensity vs radial co-ordinate) at the axial location corresponding to the peak of the wavelength averaged axial PSF and generated the 2D PSFs by rotating over 360º. Finally, these 2D PSFs for each wavelength and their average, as well as their cross-sectional PSFs, magnified views (restricted to 10 μm × 10 μm box) and the full-width-at-half-maximum (FWHM) are shown in Fig. S4 (e).

During optimization, in order to reduce computational time, the FDTD grid size of 76.92 nm was used. After optimization, we ran the FDTD simulation using the optimized geometry, keeping all other parameters the same, but with a finer FDTD grid size of 19.6 nm. The result of this "verification" step is shown in Fig. S5. Although there is slight shift in the axial PSFs in the individual wavelengths, there is negligible shift in the location of the peak of the wavelength averaged axial PSF.

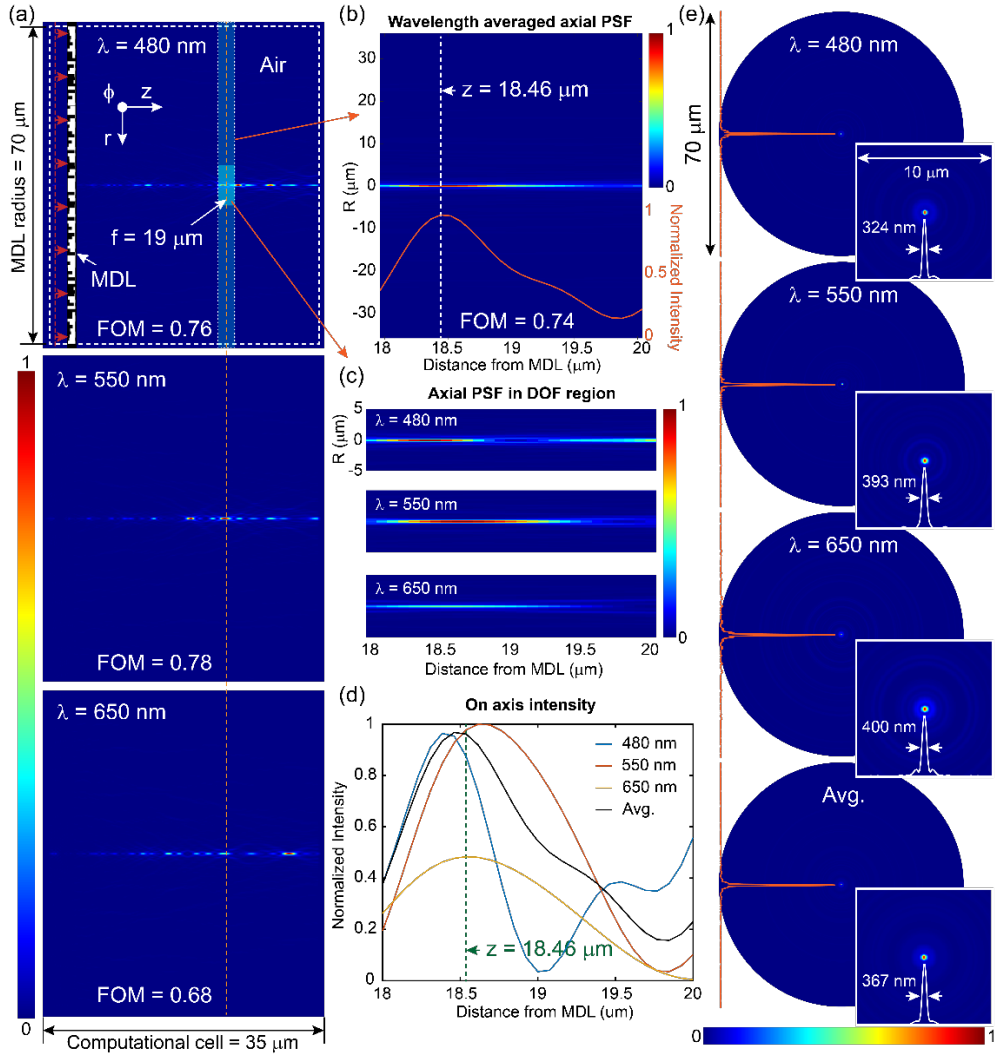

**Fig. S4**: Simulation results of the optimization process (FDTD grid size = 76.92 nm). (a) The axial PSF in the entire simulation cell for all three design wavelengths (480, 550 and 650 nm). Although half the cell is simulated, for easier data visualization, the other half is concatenated. The location of the micro-MDL is shown. The dashed box denotes the perfectly matched layer (PML). The location of the source is shown by the dashed red line and is 0.2 μm to the right of the PML left boundary. The concentration volume (radius = 35 μm, depth = 2 μm) is shown by the light blue box with dotted white line border. The focal volume (radius = 10 μm, depth = 2 μm) is shown by the bright blue box. The center of these volumes is 19 μm away from the MDL. The values of the FOM are indicated. The average of these values corresponds to the values in the plot in Fig. 1(e). (b) The wavelength averaged axial PSF in the concentration volume. The normalized intensity through the center of the axial PSF is overlaid. The peak corresponds to an on-axis location = 18.46 μm away from the micro-MDL. (c) Magnified view of the axial PSF for each individual design wavelength in the focal volume. (d) The normalized intensity through the center of the axial PSFs for the individual wavelengths as well as the average. (e) Individual 2D and cross-sectional PSFs (Normalized intensity vs radial co-ordinate) at the axial

location corresponding to the peak of the wavelength averaged axial PSF. The magnified views (restricted to 10 μm × 10 μm box) and the full-width-at-half-maximum (FWHM) are shown inset.

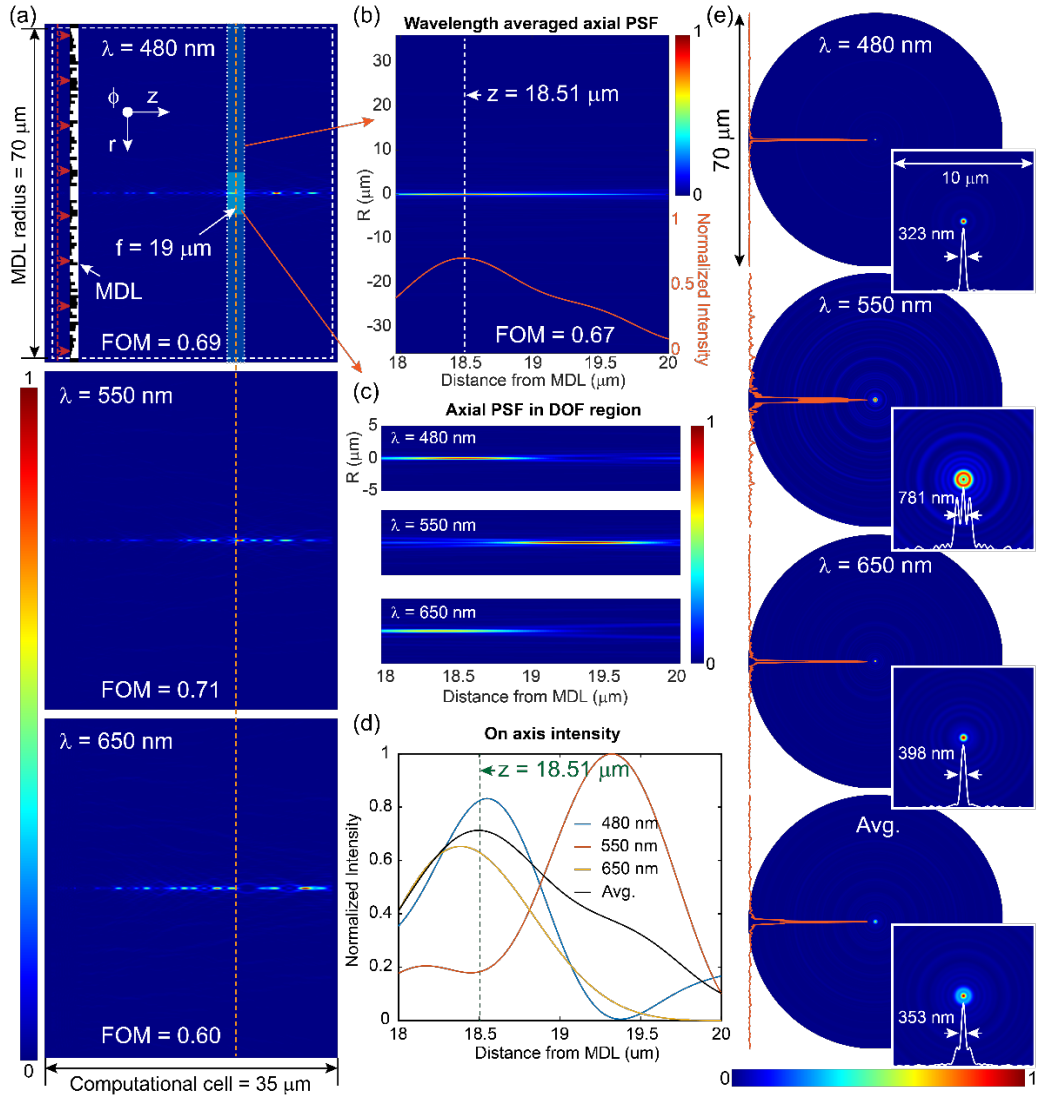

**Fig. S5**: Simulation results of the design verification process. (FDTD grid size = 19.6 nm). (a) The axial PSF in the entire simulation cell for all three design wavelengths (480, 550 and 650 nm). All parameters same as Fig. S3, except the FDTD grid resolution. (b) The wavelength averaged axial PSF in the concentration volume. The normalized intensity through the center of the axial PSF is overlaid. The peak corresponds to an on-axis location = 18.51 μm away from the micro-MDL. (c) Magnified view of the axial PSF for each individual design wavelength in the focal volume. (d) The normalized intensity through the center of the axial PSFs for the individual wavelengths as well as the average. (e) Individual 2D and cross-sectional PSFs (Normalized intensity vs radial co-ordinate) at the axial location corresponding to the peak of the wavelength averaged axial PSF. The magnified views (restricted to 10 μm × 10 μm box) and the full-width-at-half-maximum (FWHM) are shown inset.

Next, we performed the same simulation, but adding a polymer layer of thickness = 35 μm behind the MDL (FDTD grid size = 19.6 nm). The results are shown in Fig. S6. Although there is some shift in the axial PSF, most of the light is concentrated inside the 10 μm × 2μm focal volume.

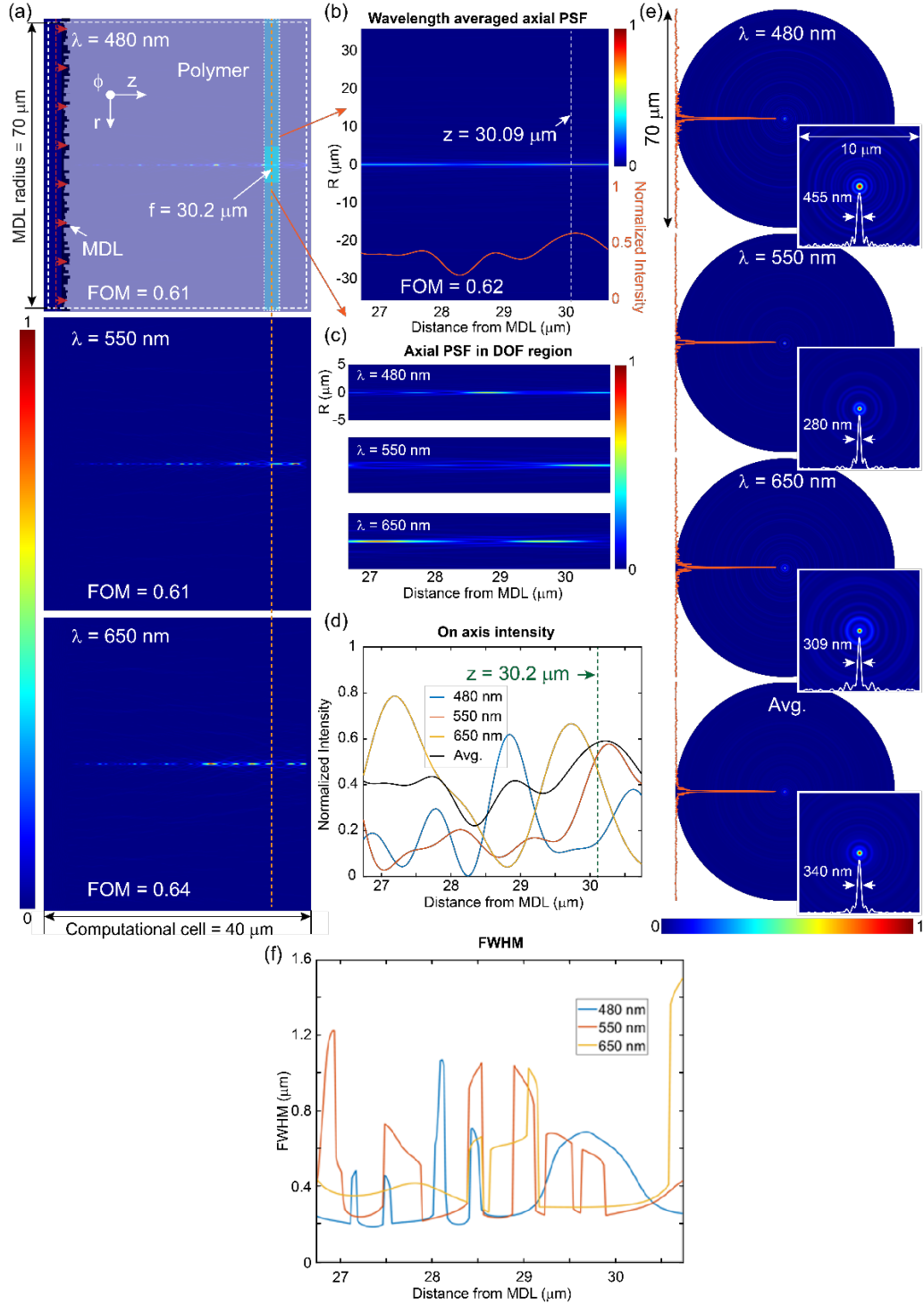

**Fig. S6**: Simulation results of the design verification process including a polymer layer of thickness = 30 μm behind the MDL (FDTD grid size = 19.6 nm). (a) The axial PSF in the entire simulation cell for all three design wavelengths (480, 550 and 650 nm). All parameters same as Fig. S3, except the FDTD grid resolution. (b) The wavelength averaged axial PSF in the concentration volume. The normalized intensity through the center of the axial PSF is overlaid. The peak corresponds to an on-axis location = 30.2 μm away from the micro-MDL. (c) Magnified view of the axial PSF for each individual design wavelength in the focal volume. (d) The normalized intensity through the center of the axial PSFs for the individual wavelengths as well as the average. (e) Individual 2D and cross-sectional PSFs (Normalized intensity vs radial co-ordinate) at the axial location corresponding to the peak of the wavelength averaged axial PSF. The magnified views (restricted to 10 μm × 10 μm box) and the full-width-at-half-maximum (FWHM) are shown inset. (f) Simulated FWHM vs distance from the MLA for design wavelengths.

## IV. DETAILS OF FLEXIBLE MLA FABRICATION

We fabricated a 200 by 200 micro-MDL array on a 50.8 mm diameter, 500 μm thick, glass substrate using grayscale lithography. The substrate was spin-coated with a positive-tone photoresist (S1813 G2 Photoresist, MICROPOSIT) [38] at 800 rpm for 60 seconds, followed by baking on a hotplate at 110°C for 2 min, followed by a one day cool down. A laser pattern generator (DWL66+, Heidelberg Instruments GmbH) patterned the micro-MDL design onto the sample. After exposure, the photoresist was developed in AZ Developer (diluted 1:1 with DI water) for 1 minute and 20 seconds. On a second fabrication, the development time was 1 min and 30 seconds. Before the lens was patterned onto the sample, a calibration sample (prepared and developed in the same way and on the same day as the actual sample previously mentioned) was exposed and developed to map the photoresist depths to corresponding laser intensities from the pattern generator.

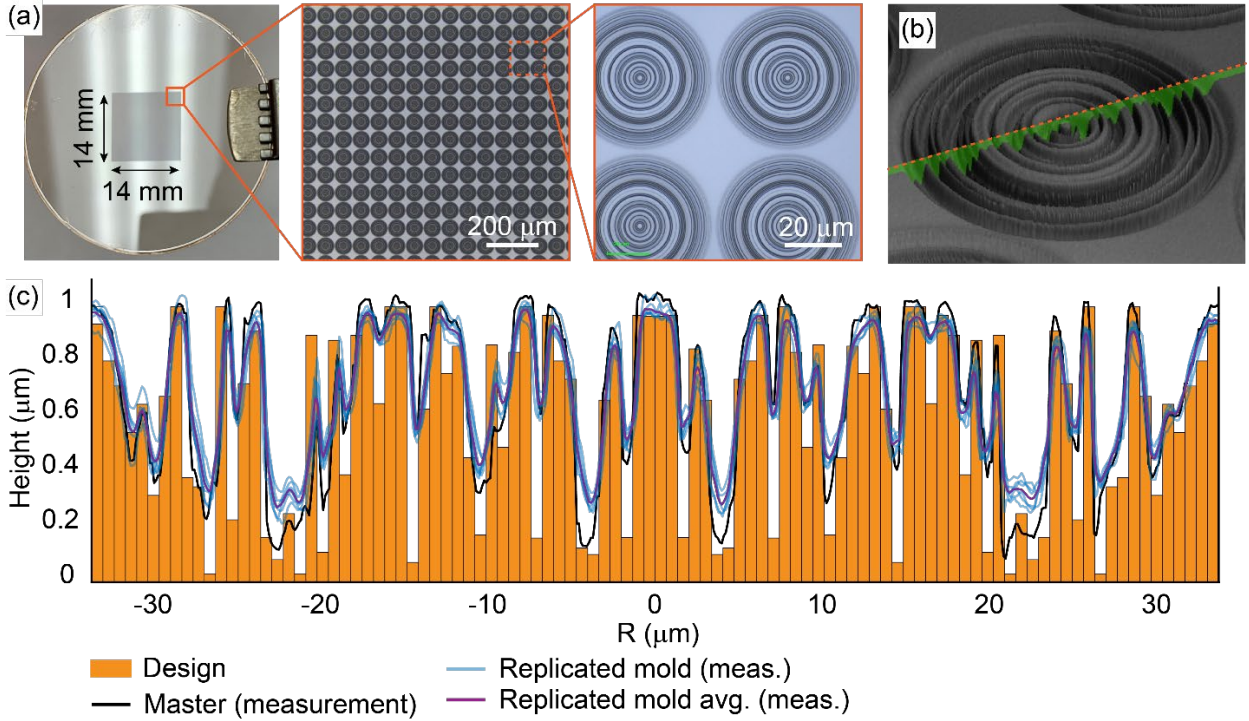

**Fig. S7**: (a) Photograph of the fabricated master showing the MLA on a glass substrate and optical micrographs of sections of the MLA at different magnifications, obtained using a confocal microscope (LEXT OLS5000, Olympus). (b) Optical micrograph of one of the micro-MDLs of the MLA in tilted view showing the cross-section used for obtaining the data for the fabrication error analysis. The orange dashed line across the MDL indicates the direction of height measurements as presented in (c). (c) Design and measured height values of the rings, across the MDL. Measurements for the master, five MDLs from the replicated molds as well as the average for the molds are presented. These measurements were obtained using the Olympus LEXT OLS5000.

Each micro-MDL in the array consisted of 50 concentric rings, as shown in Figs. 1 and 2, with a ring width of 700 nm and a maximum depth of 1 µm. The diameter of one microlens was 70 µm, and the whole 200 x 200 array was 14 mm x 14 mm. After the fabrication, the ring heights of a couple of the microlenses were measured with a confocal microscope (LEXT OLS5000, Olympus) by scanning across the diameter of one micro-MDL. Those measurements were compared with the intended heights, as shown in Fig. S6. To account for the inevitable sloping between height changes, an average of the center third of each step was compared to the intended height of that ring. The average difference between the measured and intended heights was 130 nm, with standard deviations of 176 nm. In order to simplify some of the MATLAB code involved in calculating these errors, the MATLAB function "interp" was used to interpolate a few extra values between the measurements.

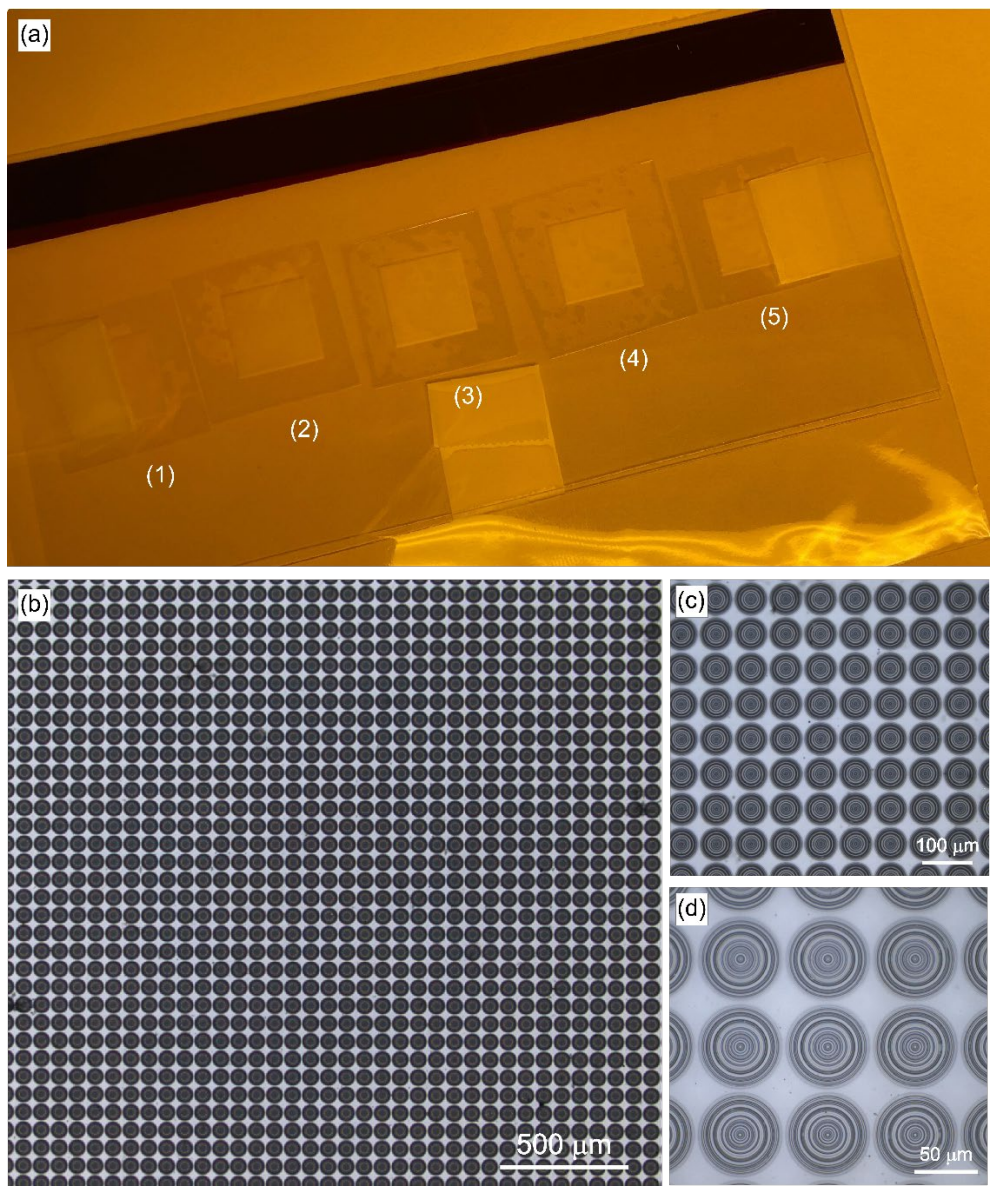

**Fig. S8**: (a) Photograph of multiple flexible MLAs produced by using nanoimprint on the master. (b-d) Optical micrographs (obtained using Olympus LEXT OLS5000) of one of the flexible MLAs at different magnifications, showing excellent replication fidelity.

The master was replicated using WaveFront Technology Inc.'s proprietary UV cast & cure process developed over the last 35 years. Using this highly developed process, surface textures with depths of between 10 nanometers up to 200 microns have been embossed. While the samples for this study were laboratory exemplars, the same materials, processes, and equipment are utilized in multiple custom-built in-house casting lines that range from a 6" wide pilot line up to a 68" wide production line capable of replicating millions of linear feet annually. Several microlenses were measured on the replicated samples with the confocal microscope (LEXT OLS5000, Olympus) by scanning across their diameters. The average difference between the measured and intended heights was 162 nm and the standard deviation was 202 nm. The heights errors and standard deviations for the mold are only about 30 nm off from its master and are presented in Fig. S7(c).

Additionally, we used an atomic force microscope (AFM, Bruker Dimension Icon) to get a more accurate metrology of the final polymer MLA device. The AFM is more accurate due to its ability of high-resolution scan and revealed that the actual average difference between the measured and intended heights was 111.4 nm and the standard deviation was 94.6 nm. This data is presented in Fig. 3(d-e).

Fig. S8 shows a photograph of multiple flexible replication samples as well as high magnification optical micrographs of some of the samples. Fig. S9 shows optical micrographs of the offset prints at different magnifications, thereby demonstrating the mass replication ability.

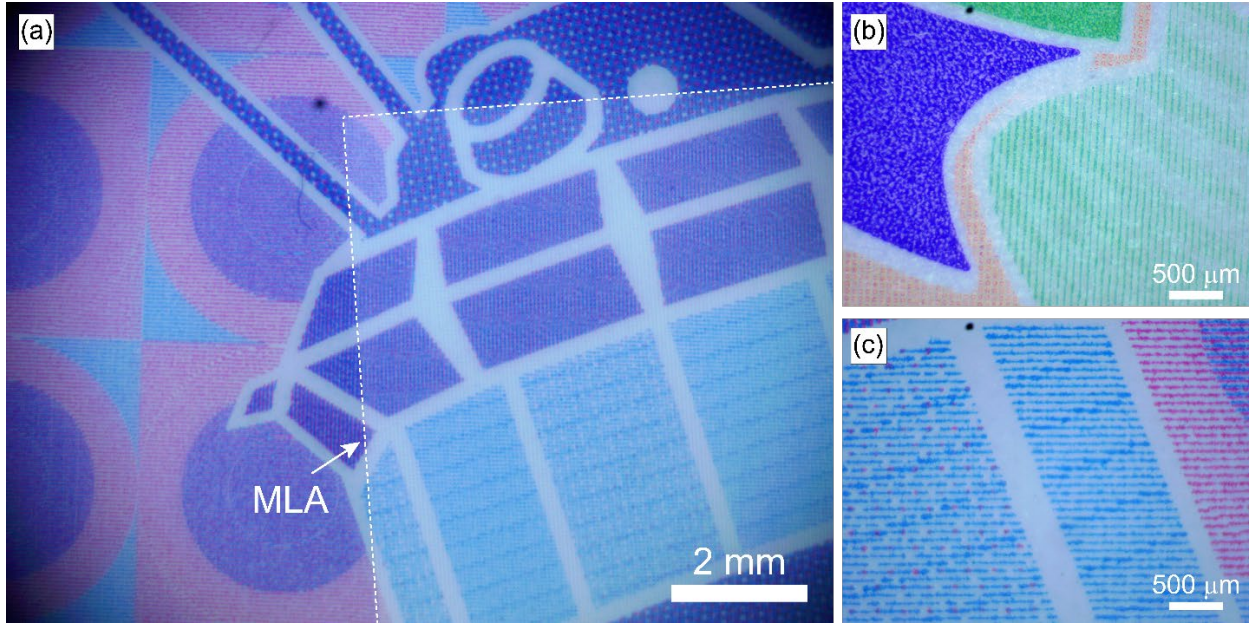

**Fig. S9**: (a) Optical micrograph of one of the "offset" prints with the MLA laid over it as shown in the dotted square. (b) Optical micrographs of other print designs at different magnifications.

## V. DETAILS OF MLA FOCUSING CHARACTERIZATION

In order to characterize the focusing performance of the MLA, we recorded its point spread function (PSF) under both narrowband and broadband illumination and for various locations on the optical axis. Fig. S10 (a) shows a schematic illustration of the optical setup used for the PSF characterization experiments. The beam from a supercontinuum source (SuperK FIANUM FIU-15, NKT Photonics) [39] coupled to a tunable filter (SuperK VARIA, NKT photonics) [40] was directed by three flat mirrors (PF10-03-P01, Thorlabs), expanded by one achromatic doublet negative lens (ACN254-050-A, Thorlabs) to fill the input aperture of parabolic mirror (CM127-012-P01, Thorlabs), expanded by it and then collimated by another parabolic mirror (CM750-500-P01, Thorlabs). This produced a 3" diameter collimated beam, i.e. planewave illumination. The expansion and collimation optics are not shown in Fig. S10 (a). The tunable filter SuperK VARIA was used to select the wavelengths of interest and corresponding bandwidth. For the narrowband measurements, the bandwidth was kept at 5 nm for each wavelength of interest.

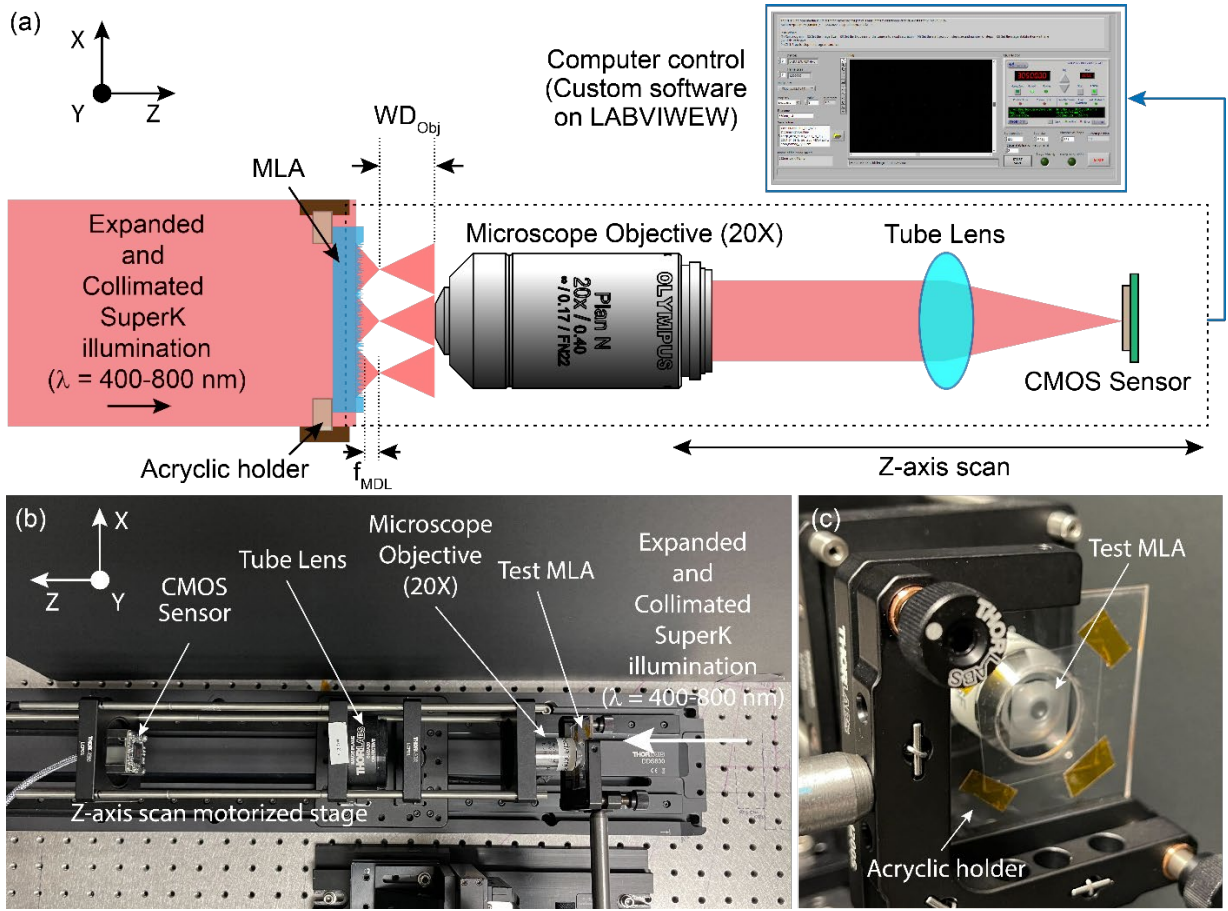

**Fig. S10**. (a) Schematic illustration of the optical setup used for the PSF characterization experiments. (b) Photograph of the same optical setup. (c) Photograph showing the flexible MLA mounted on an acrylic holder with a cut window and held down using Kapton tape at its four corners. The patterned side faces the incident illumination.

Since imaging through the polymer may introduce undesired aberrations, we positioned the MLA with the patterned side facing the objective lens. After passing through the MLA, the beam is focused at one focal length away. However, since the MLA is flexible, it creates problems for recording the PSF. To solve this, the flexible MLA was mounted on a custom-made laser cut acrylic holder with a window cut at its center. The MLA was stretched across the window and secured onto the acrylic holder using Kapton tape at its four corners to keep it flat and taut. A photograph of the optical setup is shown in Fig. S10 (b), while a photo of the flexible MLA mounted on the acrylic holder is shown in Fig. S10 (c). An iris (ID75Z, Thorlabs) was used to control the diameter of the collimated beam incident on the MLA. The MLA on glass substrate was mounted using a kinematic mount (KM200S, Thorlabs) and positioned into the path of the incident beam using a 3-axis stage fitted with micrometer actuators as shown in Fig. S9(b). Then the PSF was recorded by a setup consisting of a microscope objective (Olympus RMS20X objective, Thorlabs) [41], a tube lens (ITL200, Thorlabs) [42] and a monochromatic CMOS sensor (DMM 27UP031-ML, Imaging Source) [43]. The overall magnification of the system is 22.06X. The CMOS sensor was mounted on a motorized Z-axis scan stage (mount LCP01 and stage DDS600, Thorlabs) connected to a desktop PC running custom LabView software. The movement of the stage was synchronized with PSF frame capture on the CMOS sensor using the custom LabView software. We scanned through the optical axis with a step size of 500 nm. Cross-sections through these subsequent frames were stitched to produce the axial PSF plots shown in Fig. 4 and Fig. S12.

Fig. S11 shows the full-frame view as captured using the magnification setup and sensor. Here, the magnification is 22.06X. 9 of the focal spots from the MLA can be seen. The view is at the focal plane. The illumination in this case was λ = 550 nm with a bandwidth of 5 nm.

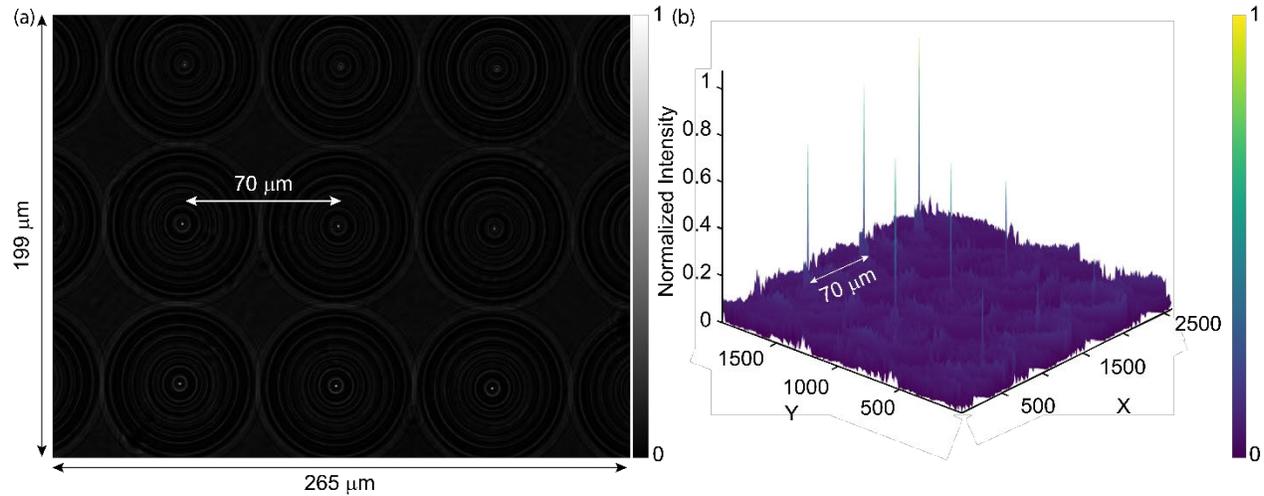

**Fig. S11.** (a) Full-frame view as captured using the magnification setup and sensor. This is the raw data from the CMOS sensor. Here, the magnification is 22.06X. 9 focal spots from the MLA can be seen. The view is at the focal plane. The illumination in this case was λ = 480 nm with a bandwidth of 5 nm. (b) Same data represented as a surface plot for easier visualization of the PSFs of the microlenses. The pitch of the PSF centers equals 70 μm, which matches with the pitch of the microlenses in the MLA.

Fig. S12 shows wavelength resolved PSF characterization results of the polymer MLA, including the axial PSF and transverse 2D PSFs.

We also simulated the performance of a Fresnel Zone Plate (FZP) with the same design parameters as the MDLs in the MLA. Since an FZP can be designed considering only one wavelength, we chose the center wavelength at 550 nm. Design parameters of FZP: design wavelength = 550 nm, f = 19 μm in air, Diameter = 70 μm, f/# = f/0.2714, and NA = 0.8789. The simulation results are presented in Fig. S13 and show strong chromatic aberrations for the non-design wavelengths as expected. As shown in Fig. S13(d) and (e), although light at 550 nm is focused close to the desired focal length, light at 480 and 650 nm is not focused at all. This analysis demonstrates the necessity to perform inverse design.

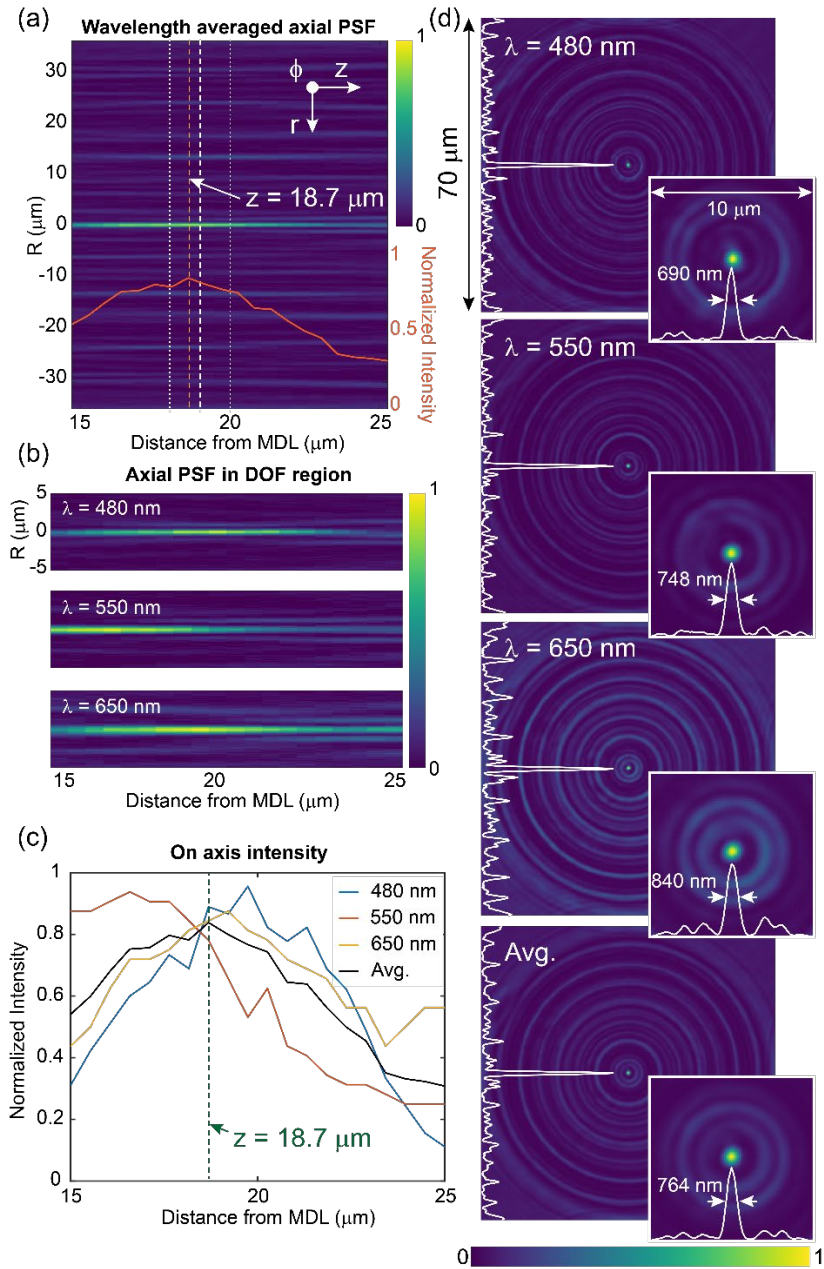

**Fig. S12**: Experimentally measured performance of the polymer MLA. (a) The wavelength averaged axial PSF in the concentration volume. The normalized intensity through the center of the axial PSF is overlaid. There is some shift in the expected location of the focal spot in the polymer. Nevertheless, a large part of the light concentration is within the +/- 1 µm DOF region as indicated by the dotted region in (a). (b) Magnified view of the axial PSF for each individual design wavelength (480, 550 and 650 nm) in the focal volume. (c) The normalized intensity through the center of the axial PSFs for the individual wavelengths as well as the average. (d) Individual 2D and cross-sectional PSFs (Normalized intensity vs radial co-ordinate) at the axial location corresponding to z = 28.5 µm of the wavelength averaged axial PSF. The magnified views (restricted to 10 µm × 10 µm box) and the full-width-at-half-maximum (FWHM) are shown inset.

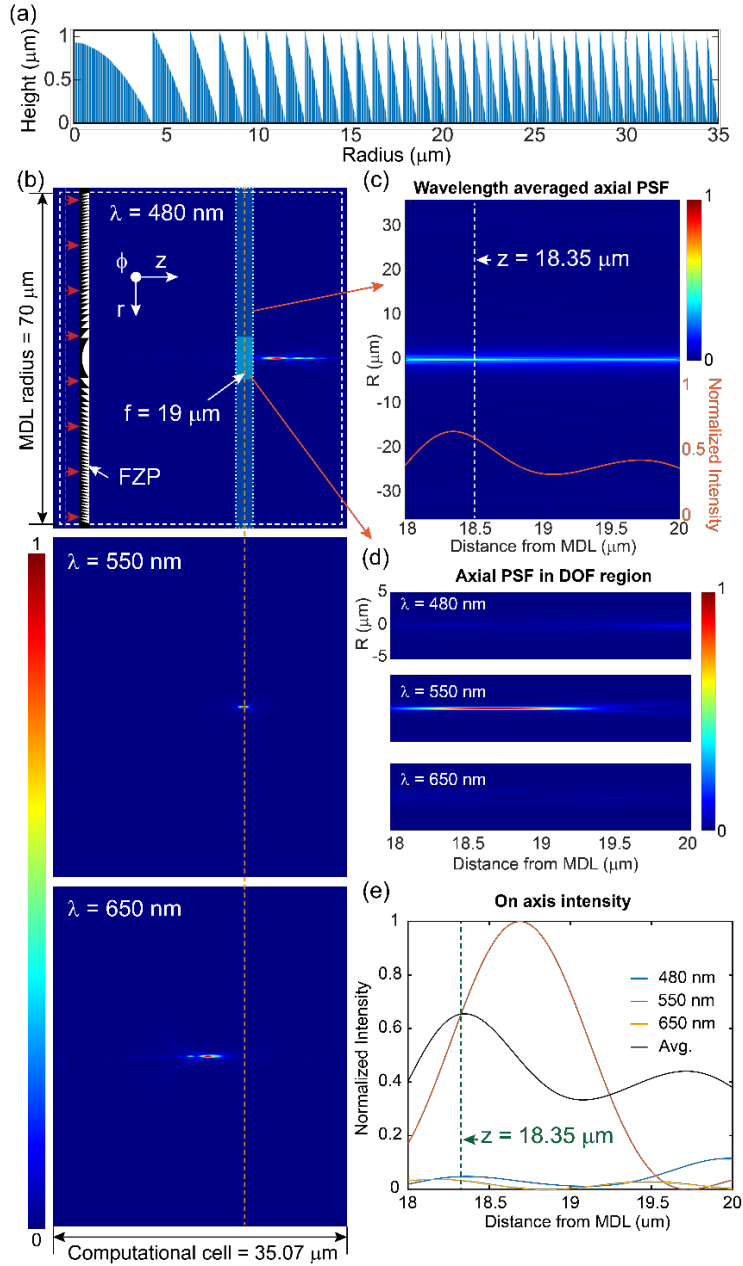

**Fig. S13**: Simulation results for a Fresnel Zone Plate (FZP) with the same design parameters as the MDLs in the MLA (design wavelength = 550 nm, f = 19 μm in air, Diameter = 70 μm, f/# = f/0.2714, and NA = 0.8789. and FDTD grid size = 19.6 nm). (a) Design of the FZP. (b) The axial PSF in the entire simulation cell for all three wavelengths (480, 550 and 650 nm). (c) The wavelength averaged axial PSF in the concentration volume. The normalized intensity through the center of the axial PSF is overlaid. The peak corresponds to an on-axis location = 18.35 μm away from the FZP. (d) Magnified view of the axial PSF for each individual design wavelength in the focal volume. Strong chromatic aberrations are observed for the non-design wavelengths 480 and 650 nm. (e) The normalized intensity through the center of the axial PSFs for the individual wavelengths as well as the average.

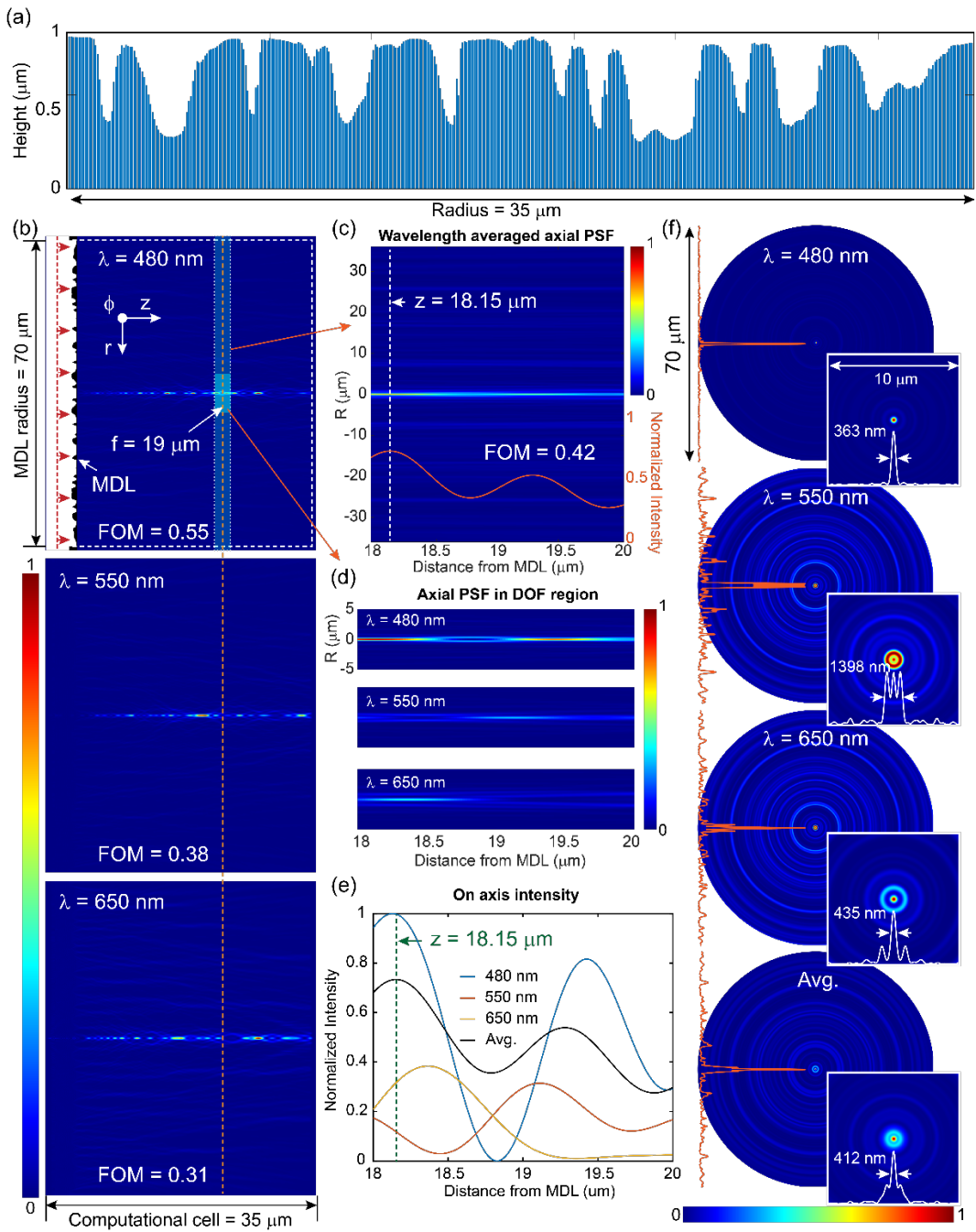

**Fig. S14**: Simulation results taking into account fabrication errors (FDTD grid size = 19.6 nm). (a) The height profile measured by the AFM as shown in Fig. 3(e), is used as the input for the simulation and thus takes into account the fabrication errors. The source is also set to have a bandwidth of 5 nm at FWHM, approximating that of the SuperK source used in the experiments. (b) The axial PSF in the entire simulation cell for all three design wavelengths (480, 550 and 650 nm). The location of the micro-MDL is shown. The patterned side faces the focal plane as is the case during experiments. (c) The wavelength averaged axial PSF in the concentration volume. The normalized intensity through the center of the axial PSF is overlaid. The peak corresponds to an on-axis location = 18.15

μm away from the micro-MDL. (d) Magnified view of the axial PSF for each individual design wavelength in the focal volume. (e) The normalized intensity through the center of the axial PSFs for the individual wavelengths as well as the average. (f) Individual 2D and cross-sectional PSFs (Normalized intensity vs radial co-ordinate) at the axial location corresponding to the peak of the wavelength averaged axial PSF. The magnified views (restricted to 10 μm × 10 μm box) and the full-width-at-half-maximum (FWHM) are shown inset.

## VI. PERFORMANCE CHARACTERIZATION OF 3D INTEGRAL IMAGES

Fig. 5(a) shows integration of the MLA with an "offset" security print to produce integral image. In most cases, ambient light was used to back-illuminate this combination. The 3D integral images could be easily viewed with naked eye. In order to record the performance, we capture photos and videos using an iPhone 12 Pro Max camera.

## VII. DESCRIPTION OF SUPPLEMENTARY VIDEOS

Images and videos of the integral images were also recorded using an iPhone 12 Pro Max camera under various illumination conditions and summarized in Figs. 4(c-e).

Visualization 1 shows one of the offset prints without the MLA, exhibiting no 3D integral imaging.

Visualization 2 shows transparent prints integrated with the MLA exhibiting 3D integral imaging under different illumination conditions. The illuminations were: (a) All room lights off, only one single white LED flashlight turned on. (b) All room lights on. Print held up against ceiling fluorescent light. (c) Ambient sunlight, captured outside around 5 pm, on November 08, 2023 at Salt Lake City, Utah, USA.

Visualization 3 shows the same print design but in (a) transparent and (b) opaque versions, exhibiting integral imaging in combination with the MLA.

Visualization 4 and 5 show the simulated 3D animation of the integral imaging that is expected to be achieved by the prints and the MLA.

We note that the experimental images agree well with the simulations (Fig. 4(b)), confirming the accuracy of our fabrication processes.

Visualization 6 shows the transparent prints integrated with the MLA and fixed using Kapton tape at the edges, exhibiting 3D integral imaging and also the effect of moving and rotating the MLA with respect to the print to generate different Moiré patterns that do not resemble the simulated integral images.

## REFERENCES

1. S. Shrestha, A. Overvig, M. Lu, A. Stein, N. Yu, "Broadband achromatic dielectric metalenses" Light: Sci. Appl. 2018, 7, 85.
2. L. Zhang, J. Ding, H. Zheng, S. An, H. Lin, B. Zheng, Q. Du, G. Yin, J. Michon, Y. Zhang, and Z. Fang, "Ultra-thin high-efficiency mid-infrared transmissive Huygens meta-optics." Nat. comm., 9(1), p.1481 (2018).
3. S. Zhang, A. Soibel, S.A. Keo, D. Wilson, S. Rafol, D.Z. Ting, A. She, S.D. Gunapala, and F. Capasso, "Solid-Immersion Metalenses for Infrared Focal Plane Arrays." Appl. Phys. Lett. 113, 111104 (2018).
4. A. Majumder, M. Meem, R. Stewart, and R. Menon, "Broadband point-spread function engineering via a freeform diffractive microlens array," Opt. Express 30, 1967-1975 (2022).
5. J. Hu, C.H. Liu, X. Ren, L.J. Lauhon, and T.W. Odom, "Plasmonic lattice lenses for multiwavelength achromatic focusing." ACS nano, 10(11), pp.10275-10282, (2016).

6. H. Zuo, D.Y. Choi, X. Gai, P. Ma, L. Xu, D.N. Neshev, B. Zhang, and B. Luther-Davies, "High-Efficiency All-Dielectric Metalenses for Mid-Infrared Imaging." Adv. Opt. Mat., 5(23), p.1700585, (2017).

7. S. Colburn, A. Zhan and A. Majµmdar, "Metasurface optics for full-color computational imaging," Science Advances, 4 (2), (2018).

8. M. Meem, S. Banerji, A. Majumder, F. Guevara Vasquez, B. Sensale-Rodriguez and R. Menon, "Broadband lightweight flat lenses for longwave-infrared imaging," Proc. Natl. Acad. Sci. 116 (43), 21375-21378 (2019).

9. H. C. Wang, C.H. Chu, P.C. Wu, H. H. Hsiao, H.J. Wu, J.-W. Chen, W.H. Lee, Y. C. Lai, Y. W Huang, M.L. Tseng, S. W. Chang, D.P. Tsai, "Ultrathin Planar Cavity Metasurfaces" Small, 14, 1703920 (2018).

10. M. Meem, S. Banerji, A, Majumder, C. Pies, T. Oberbiermann, B. Sensale-Rodriguez & R. Menon, "Inverse-designed flat lens for imaging in the visible & near-infrared with diameter > 3mm and NA=0.3," Appl. Phys. Lett. 117(4), 041101 (2020).

11. S. Wang, P.C. Wu, V.C. Su, Y.C. Lai, C.H. Chu, J.W. Chen, S.H. Lu, J. Chen, B. Xu, C.H. Kuan, and T. Li, "Broadband achromatic optical metasurface devices." Nat. comm., 8(1), p.187 (2017).

12. F. Balli, M. Sultan, S. K. Lami, et al., "A hybrid achromatic metalens," Nat Commun 11, 3892 (2020).

13. Lin, R.J., Su, VC., Wang, S. et al. Achromatic metalens array for full-colour light-field imaging. Nat. Nanotechnol. 14, 227–231 (2019).

14. M. Khorasaninejad, Z. Shi, A. Y. Zhu, W. T. Chen, V. Sanjeev, A. Zaidi, and F. Capasso, "Achromatic Metalens over 60 nm Bandwidth in the Visible and Metalens with Reverse Chromatic Dispersion," Nano Letters, 17 (3), 1819-1824 (2017).

15. W.T. Chen, A.Y. Zhu, V. Sanjeev, M. Khorasaninejad, Z. Shi, E. Lee, and F. Capasso, "A broadband achromatic metalens for focusing and imaging in the visible." Nat. Nano., 13(3), p.220 (2018).

16. Chen, W.T., Zhu, A.Y., Sisler, J. et al. A broadband achromatic polarization-insensitive metalens consisting of anisotropic nanostructures. Nat Commun 10, 355 (2019).

17. S. Banerji, M. Meem, A. Majumder, C. Dvonch, B. Sensale-Rodriguez, and R. Menon, "Single flat lens enabling imaging in the short-wave infra-red (SWIR) band," OSA Continuµm 2, 2968-2974 (2019).

18. M. Ye, V. Ray, and Y.S. Yi, "Achromatic Flat Subwavelength Grating Lens Over Whole Visible Bandwidths." IEEE Phot. Tech. Lett., 30(10), pp.955-958 (2018).

19. Ndao, A., Hsu, L., Ha, J. et al. Octave bandwidth photonic fishnet-achromatic-metalens. Nat Commun 11, 3205 (2020).

20. S. Wang, P.C. Wu, V.C. Su, Y.C. Lai, M.K. Chen, H.Y. Kuo, B.H. Chen, Y.H. Chen, T.T. Huang, J.H. Wang, and R.M. Lin, "A broadband achromatic metalens in the visible." Nat. Nano., 13(3), p.227 (2018).

21. S. Wang, C. Zhou, Z. Liu, and H. Li, "Design and analysis of broadband diffractive optical element for achromatic focusing." In Holography, Diffractive Optics, and Applications VII (Vol. 10022, p. 100221J) (2016).

22. Fan, ZB., Qiu, HY., Zhang, HL. *et al.* A broadband achromatic metalens array for integral imaging in the visible. *Light Sci Appl* **8,** 67 (2019).

23. S. Banerji, M. Meem, A. Majumder, B. Sensale-Rodriguez, and R. Menon, "Super-resolution imaging with an achromatic multi-level diffractive microlens array," Opt. Lett. 45, 6158-6161 (2020).

24. M. Meem, A. Majumder, and R. Menon, "Free-form broadband flat lenses for visible imaging," OSA Continuµm 4, 491-497 (2021).

25. M. Khorasaninejad, F. Aieta, P. Kanhaiya, M.A. Kats, P. Genevet, D. Rousso, and F. Capasso, "Achromatic metasurface lens at telecommunication wavelengths." Nano lett. 15(8), pp.5358-5362 (2015).

26. N. Mohammad, M. Meem, B. Shen, P. Wang and R. Menon, "Broadband imaging with one planar diffractive lens," Sci. Rep. 8 2799 (2018).

27. M. Meem, A. Majumder and R. Menon, "Full-color video and still imaging using two flat lenses," Opt. Exp. 26(21) 26866-26871 (2018).

28. S. Banerji, M. Meem, A, Majumder, F. Vasquez-Guevara, B. Sensale-Rodriguez & R. Menon, "Ultra-thin near infrared camera enabled by a flat multi-level diffractive lens," Opt. Lett. 44(22) 5450-5452 (2019).

29. M. Meem, S. Banerji, A, Majumder, C. Pies, T. Oberbiermann, B. Sensale-Rodriguez & R. Menon, "Large-area, high-NA multi-level diffractive lens via inverse design," Optica 7(3) 252-253 (2020).

30. M. Meem, S. Banerji, A, Majumder, J. C. Garcia, O. Kigner, P. Hon, B. Sensale-Rodriguez & R. Menon, "Imaging from the visible to the longwave infrared via an inverse-designed flat lens," Opt. Exp. 29(13) 20715-20723 (2021).

31. T. M. Hayward, et al., "Multilevel diffractive lens in the MWIR with extended depth-of-focus and wide field-of-view," Opt. Exp. 31(10), 15384-15391 (2023).

32. D. Lin, T. M. Hayward, W. Jia, A. Majumder, B. Sensale-Rodriguez, and R. Menon, "Inverse-designed multi-level diffractive doublet for wide field-of-view imaging," ACS Photonics 10, 8, 2661–2669 (2023).

33. A. Taflove and S.C. Hagness, "Computational Electrodynamics: The Finite-Difference Time-Domain Method", Artech: Norwood, 3rd Edition, MA, 2005.

34. A. Oskooi, D. Roundy, M. Ibanescu, P. Bermel, J.D. Joannopoulos, and S.G. Johnson, "MEEP: A flexible free-software package for electromagnetic simulations by the FDTD method," Computer Physics Communications, Vol. 181, pp. 687-702 (2010).

35. MEEP Cylindrical coordinates tutorial: https://meep.readthedocs.io/en/latest/Scheme_Tutorials/Cylindrical_Coordinates/#:~:text=Meep%20s upports%20the%20simulation%20of,if%20there%20is%20sufficient%20symmetry.

36. P. Wang, N. Mohammad and R. Menon, "Chromatic-aberration-corrected diffractive lenses for ultra-broadband focusing," Sci. Rep., 6, 21545 (2016).

37. S. Banerji and B. Sensale-Rodriguez, "A computational design framework for efficient, fabrication error-tolerant, planar THz diffractive optical elements," Sci. Rep., 9, 5801 (2019).

38. Datasheet for S1813, Shipley: https://amolf.nl/wp-content/uploads/2016/09/datasheets_S1800.pdf

39. Datasheet for SuperK Fianum, NKT Photonics: https://www.nktphotonics.com/product-manuals-and-documentation/

40. Datasheet for SuperK VARIA, NKT photonics: https://www.nktphotonics.com/product-manuals-and-documentation/

41. Datasheet for Olympus PLN 20X objective, Thorlabs: https://www.thorlabs.com/thorproduct.cfm?partnumber=RMS20X

42. Datasheet for ITL200, Thorlabs, Thorlabs: https://www.thorlabs.com/thorproduct.cfm?partnumber=ITL200

43. Datasheet for DMM 27UP031-ML, Imaging Source: https://dl-gui.theimagingsource.com/en_US/6a01b323-02fa-59da-815a-b81ff1d6feb7/